\documentclass{article}
\usepackage{graphicx}
\usepackage{amssymb}

\begin{document}



\begin{center}

\begin{large}
{\bf Perspectives for the detection and measurement of Supersymmetry in 
the focus point region of mSUGRA models with the ATLAS detector at LHC}
\end{large}

\vspace{0.5cm}

U. De Sanctis, T. Lari, S. Montesano, C. Troncon

{\em Universit\`a di Milano - Dipartimento di Fisica and 
Istituto Nazionale di Fisica Nucleare - Sezione di Milano, 
Via Celoria 16, I-20133 Milan, Italy}

\end{center}

\vspace{0.5cm}

\begin{abstract}
This paper discusses the ATLAS potential to study Supersymmetry  
in the ``focus point'' region of the parameter space of mSUGRA models. 
The potential to discover a deviation from Standard Model expectations 
with the first few $\mbox{fb}^{-1}$ of LHC data was studied using the 
parametrized simulation of the ATLAS detector. Several signatures were 
considered, involving hard jets, large missing energy, and 
either $b$-tagged jets, opposite-sign isolated electron or 
muon pairs, or top quarks reconstructed exploiting 
their fully hadronic decays.  
With only 1~$\mbox{fb}^{-1}$ of data each of these signatures may allow 
observing  
an excess of events over Standard Model expectation with a statistical 
significance exceeding 5 standard deviations.
Furthermore, each of the two invariant mass distributions of the two 
leptons produced by the
$\tilde \chi^0_3 \rightarrow \tilde \chi^0_1 l^+ l^-$ and the 
$\tilde \chi^0_2 \rightarrow \tilde \chi^0_1 l^+ l^-$ three-body decays
has a kinematic endpoint which 
measures the difference between the masses of the 
parent and daughter neutralino. 
An analytical expression was derived 
for the shape of this distribution and was used to fit the 
simulated LHC data. A measurement of the  
$\tilde \chi^0_2 - \tilde \chi^0_1$ and  
$\tilde \chi^0_3 - \tilde \chi^0_1$ mass differences
was obtained and this information
was used to constrain the MSSM parameter space.
\end{abstract}

\section{Introduction}

One of the best motivated extensions of the Standard Model is 
the Minimal SuperSymmetric Model (MSSM)~\cite{SUSY}. Because of the large 
number of free parameters of the general MSSM, the studies in preparation
for the analysis of LHC data are mostly performed in a more constrained 
framework, obtained making some assumptions about the breaking mechanism of 
Supersymmetry. Most studies are performed in the minimal SUGRA 
framework~\cite{SUSY},
which has five free parameters: the common mass $m_0$ of scalar particles 
at the grand-unification energy scale, the common fermion mass $m_{1/2}$, 
the common trilinear coupling $A_0$, the sign of the Higgsino mass parameter 
$\mu$ and the ratio $\tan \beta$
between the vacuum expectation values of the two Higgs doublets. 

Constraints are provided by searches made by experiments at accelerators 
(in particular LEP~\cite{LEP}) and by the requirements that the radiative  
electroweak symmetry breaking is consistent with the Standard Model. 
A strong point of Supersymmetry is that in case of exact R-parity 
conservation the lightest SUSY particle (LSP) is 
stable and can thus provide a candidate for Dark Matter. 
Because  of cosmological considerations the 
LSP must be neutral and weakly interacting and in mSUGRA the suitable 
candidate is the lightest neutralino $\tilde \chi^0_1$. It is therefore 
natural to apply the additional constraint that the neutralino relic density 
$\Omega_{\tilde \chi}$ in the present 
universe should be compatible with the density of non-baryonic 
Dark Matter, which is $\Omega_{DM} h^2 = 0.105^{+0.007}_{-0.013}$~\cite{WMAP}. 
If there are other contributions to 
the Dark Matter one may have $\Omega_{\tilde \chi} < \Omega_{DM}$. 

In most of the 
mSUGRA parameter space, however, the neutralino relic density is larger
than $\Omega_{DM}$~\cite{Ell03}. An acceptable value of relic density 
is obtained only in particular regions of the parameter space, noticeably:

\begin{itemize}

\item in a region with a relatively low value of the SUSY mass scale 
({\em bulk region}).

\item for $m_{1/2} >> m_0$, when the mass of the scalar $\tau$ 
is close to the 
mass of the lightest neutralino, so that $\tilde \chi \tilde \tau$ 
annihilation 
in the early universe reduces the relic density ($\tilde \tau$ 
{\em co-annihilation region}).

\item for large value of $\tan \beta$, there is a funnel in the parameter
space where the mass of the pseudo-scalar Higgs boson is nearly twice the one 
of the neutralino, enhancing the $\tilde \chi \tilde \chi$ annihilation 
cross section ({\em Higgs funnel region}).

\item for $m_{1/2} << m_0$ a region 
exists~\cite{Focuspoint}, where the lightest neutralino 
has a significant Higgsino component, 
enhancing the $\tilde \chi \tilde \chi$ annihilation cross section 
({\em focus point region}).

\end{itemize}

In this paper a study of the ATLAS potential to discover and study 
Supersymmetry in the focus point scenario is presented.

The paper is organized as follows. In Section~\ref{sec2} a scan of the 
minimal SUGRA parameter space is performed to map the focus point 
region with an acceptable relic density. The theoretical uncertainties on 
the SUSY mass spectrum and the variation of the masses across the 
focus point parameter space are discussed. Finally, a benchmark point 
is selected for more detailed studies. For this benchmark, the production 
of Supersymmetric particles is dominated by the electroweak production of 
neutralinos and charginos, which is not easily observable in a hadronic 
collider, and by the pair production of gluinos, followed by the decay of 
each gluino into (mostly) third-generation quarks and a neutralino or chargino.
The gluino pair production results in events with hard jets, including 
4 $b$-jets per event, large missing energy from the lightest neutralinos 
which escape detection, and possibly isolated leptons from neutralino 
and chargino decays. These signatures are used in the studies reported in 
Section~\ref{sec3} to~\ref{sec5}.

In Section~\ref{sec3} the performance 
of inclusive search strategies based on the presence of hard jets, 
large missing energy, and $b$-tagged jets to discriminate the SUSY signal 
from the Standard Model background is studied. 

In Section~\ref{sec4} the 
reconstruction of the kinematic edge of the invariant mass distribution
of the two leptons from the decay $\tilde \chi^0_n \rightarrow \tilde 
\chi^0_1 l^+ l^-$ is discussed. The presence of isolated lepton pairs 
(in addition to hard jets and missing energy) is a promising discovery 
channel with the first few $\mbox{fb}^{-1}$ of LHC data; with a larger 
dataset, the kinematic endpoints of the two-lepton invariant 
mass distribution allow the measurement of two constraints on the 
masses of the three lightest neutralinos. To this purpose, an analytical 
expression for the shape of the invariant mass distribution of the 
lepton pairs arising from the three-body leptonic decay of the 
neutralinos has been derived, in the hypothesis of heavy scalar 
masses (which is a suitable approximation for the focus point). The resulting 
function was used in the fit of the dilepton invariant mass 
distribution obtained 
with simulated data, obtaining a measurement of the 
$\tilde \chi^0_2 - \tilde \chi^0_1$ and 
$\tilde \chi^0_3 - \tilde \chi^0_1$ mass differences.

In Section~\ref{sec5} the reconstruction of the gluino 
decays is discussed. To this purpose, the top quark fully-hadronic 
decays are reconstructed; the presence of an excess of $tb$ and 
$tt$ pairs is a possible discovery channel while their invariant 
mass places constraints on the gluino mass scale. 

Finally, Section ~\ref{sec6} investigates 
which constraints one can put on the Supersymmetry parameters 
from the measurements of the two neutralino mass differences. 

\section{Scans of mSUGRA parameter space}
\label{sec2}

In order to find the regions of the mSUGRA parameter space which have a 
relic density compatible with cosmological measurements, the neutralino
relic density was computed with micrOMEGAs 1.31~\cite{mOMEGAs}, interfaced 
with either ISAJET 7.71~\cite{ISA} or SOFTSUSY 1.9~\cite{SOFT} for the  
solution of the Renormalization Group Equations (RGE) to compute 
the Supersymmetry mass spectrum at the weak scale.

The two spectrum calculators both consider 2-loop radiative
corrections in computing the running of masses and couplings
between the electroweak and the unification energy scales. 
They differ in the implementation of these corrections, 
however, and the difference between their results  
reflects the uncertainties on the contributions from higher 
order radiative corrections. A detailed comparison between the RGE
calculators can be found in ref.~\cite{All03}, while the 
uncertainties on the resulting predictions of relic density are 
discussed in ref.~\cite{All04}. Over most of parameter 
space, the mass spectrum is  predicted with an uncertainty of less than 
1\%, but in the focus-point region larger differences are 
found: in this region of parameter space the results are 
particularly sensitive to the value of the top Yukawa couplings 
and the uncertainties from higher order corrections are significant.

\begin{figure}[!p]
\begin{center}
\includegraphics[width=10cm]{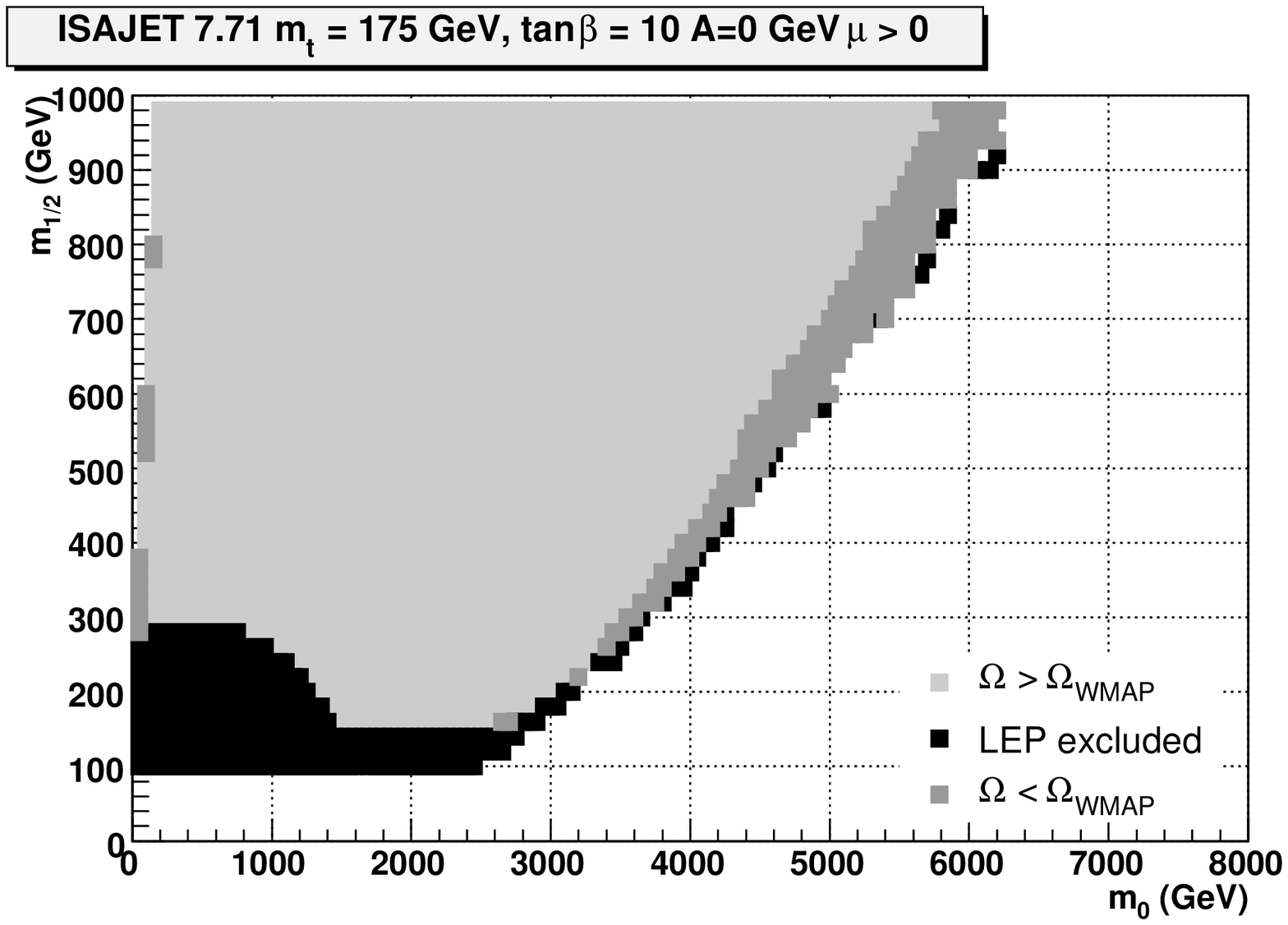}
\end{center} 
\caption{\label{scan1} The picture shows the regions of the 
$(m_0,m_{1/2})$ mSUGRA plane 
which have a neutralino relic density compatible with cosmological 
measurements in dark grey. The black
regions are excluded by LEP. The light grey regions have a 
neutralino relic density which exceeds cosmological measurements. 
White regions are theoretically excluded. The values of 
$\tan \beta = 10$, $A_0 = 0$, a positive $\mu$, and a top mass
of 175~GeV were used. The RGE were solved using ISAJET. 
}
\end{figure}

In Fig.~\ref{scan1} a scan of the $(m_0,m_{1/2})$ plane performed with 
ISAJET+micrOMEGAs is presented, for 
fixed values of $\tan \beta = 10$, $A_0 = 0$, and positive $\mu$. A top 
mass of 175~GeV was used. The dark grey region on the left is the 
scalar tau co-annihilation strip, while that on the right is the focus point 
region with $\Omega_{\tilde \chi} < \Omega_{DM}$. 

The latter is found at large value of $m_0 > 3$~TeV, hence in this scenario 
the scalar particles 
are very heavy, near or beyond the sensitivity limit of LHC searches.
Since $m_{1/2} << m_0$, the gauginos (chargino and neutralino) and gluino 
states are much lighter. In this scenario the SUSY production cross 
section at the LHC 
is thus dominated by gaugino and gluino pair production. 

\begin{figure}[!p]
\begin{center}
\includegraphics[width=7cm]{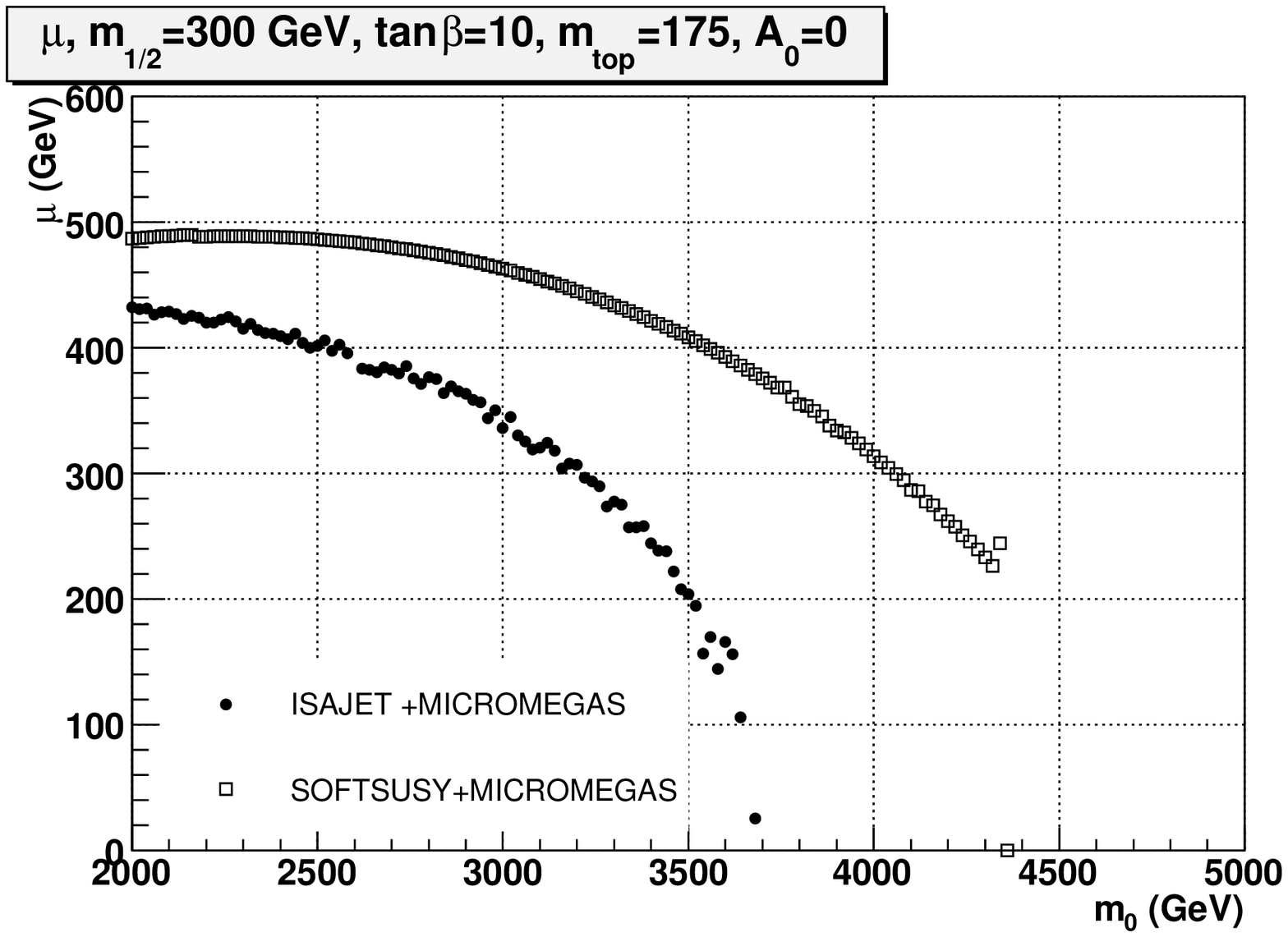}
\includegraphics[width=7cm]{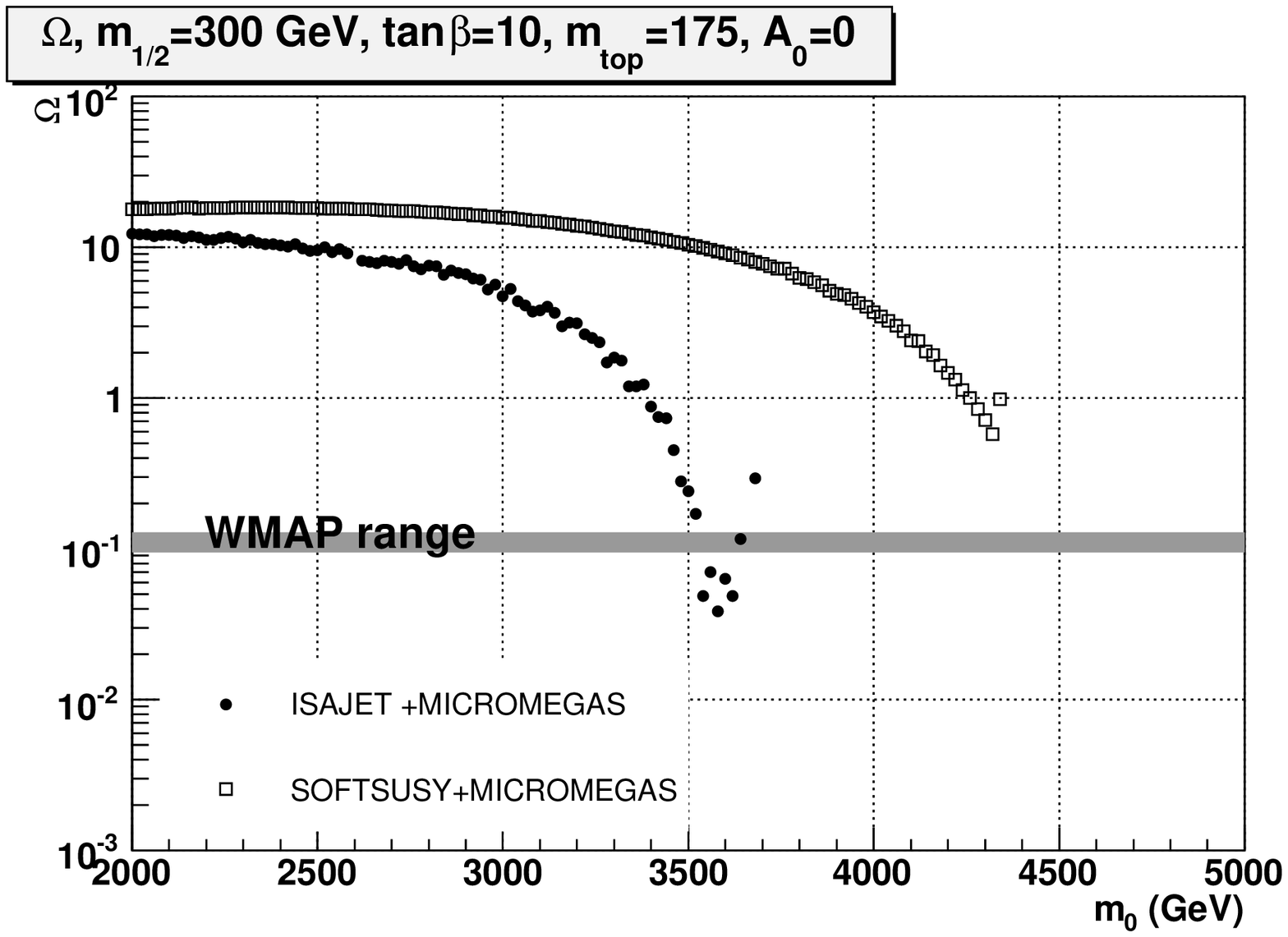}
\end{center} 
\caption{\label{scan2} Left plot: dependence of the Higgsino mass term  
$\mu$ on the mSUGRA common scalar mass $m_0$, for $m_{1/2} = 300$~GeV,
$\tan \beta = 10$, $A_0 = 0$, a positive $\mu$, and a top mass
of 175~GeV. The circles are obtained using ISAJET to solve the RGEs, 
the open squares using SOFTSUSY. Right plot: dependence of the 
neutralino relic density on $m_0$. 
}
\end{figure}

It is instructive to consider the variation of the relic density and 
the Higgsino mass term $\mu$ along a line in the  $(m_0,m_{1/2})$ 
plane at fixed $m_{1/2}$. This is shown in Fig.~\ref{scan2} for 
$m_{1/2} = 300$~GeV. The left plot reports the dependence of $\mu$ 
on $m_0$. When the value of $\mu$ drops, the lightest neutralino 
acquires a significant Higgsino component and the relic density 
decreases (as shown in the right plot): this is the focus point 
region. The picture shows that ISAJET and SOFTSUSY 
are in reasonable agreement for low values of $m_0$ but their 
predictions diverge as the scalar mass is increased, and they 
find the drop of $\mu$ at different values of $m_0$.
In addition, the value computed by SOFTSUSY never gets low enough 
to result in an acceptable value of the relic density for this 
particular choice of the parameters. 
These uncertainties make it difficult to decide whether a 
given SUSY mass spectrum is really consistent with the mass and coupling 
unification at the high scale assumed by mSUGRA. 
However, this does not prevent us to select a benchmark point 
with an interesting phenomenology, compatible with accelerator and 
cosmological constraints, and study the ATLAS potential to study this model. 

The gluino mass increases with $m_{1/2}$. It is about 800~GeV for 
$m_{1/2} = 300$~GeV, at the bottom of the focus point strip 
allowed by cosmological constraints and accelerator searches. 
This value corresponds to a cross section for gluino pair production at 
the LHC of about 1~pb.
The gluino decays to $\tilde \chi qq$ (a chargino or neutralino, 
and two quarks) followed by the cascade decays of the chargino (or neutralino) 
into the lightest neutralino. These events have the classical 
mSUGRA signature of hard jets and missing energy and can be discriminated 
from the Standard Model background as will be shown in Section~\ref{sec3}. 

As one moves upward inside the focus point
strip, the gluino mass increases, and it 
reaches the value of 2~TeV for $m_{1/2} \sim 900$~GeV. This is expected 
to be roughly the limit of the ATLAS discovery potential~\cite{ATDR}.
The mass of the lightest neutralino and chargino states is much smaller than 
the gluino and squark masses, and the cross section for 
$\tilde \chi \tilde \chi$ production at the LHC can be larger than the 
cross section for gluino pair production.  
Since the gauginos are very light, however, their decay does not provide
a signature with hard jets and missing energy, which would allow 
discrimination against the Standard Model background. The possibility to 
detect this signal through multi-lepton signatures is under study, and is 
outside the scope of this note. 

\begin{figure}[!p]
\begin{center}
\includegraphics[width=7cm]{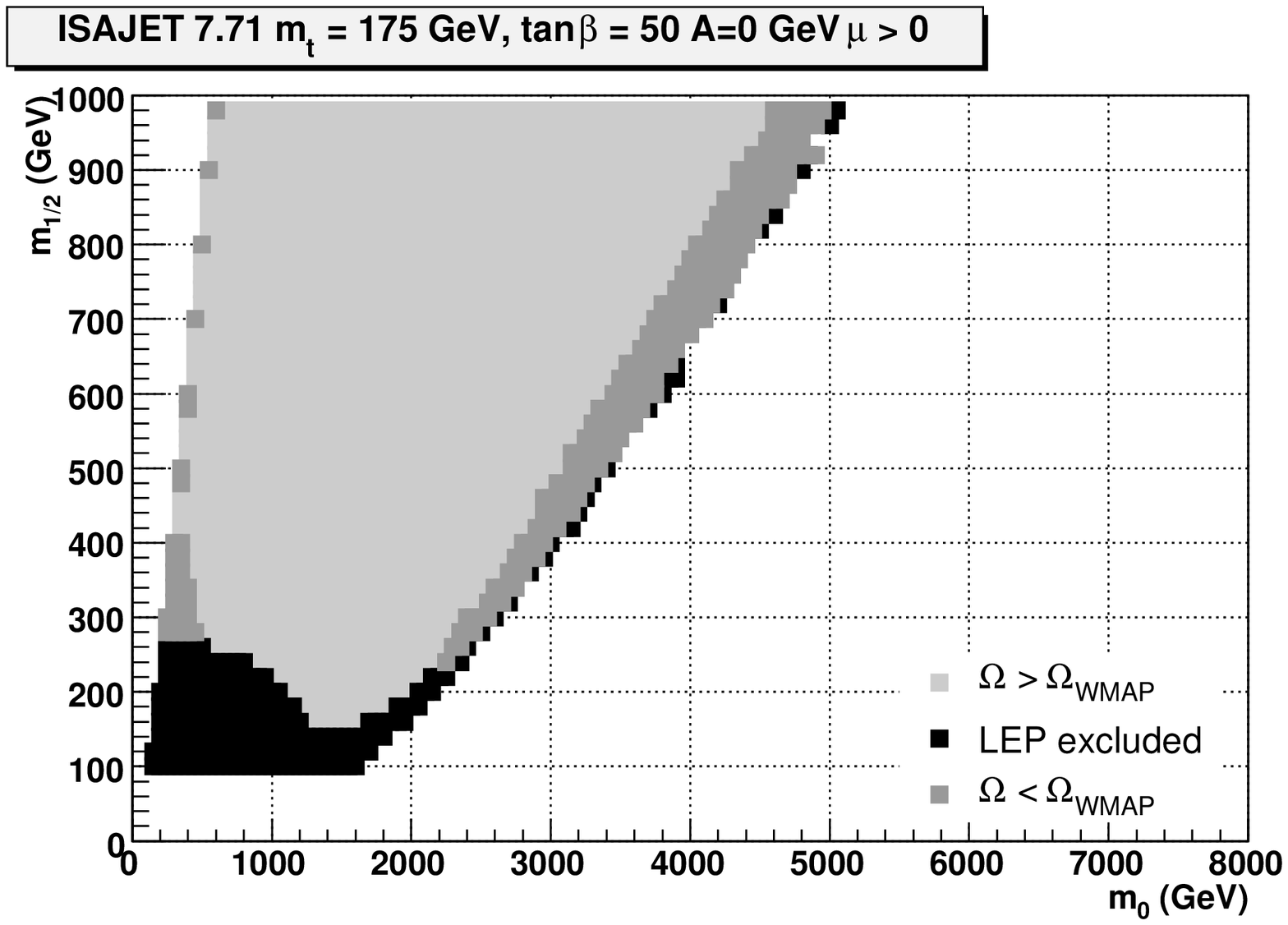}
\includegraphics[width=7cm]{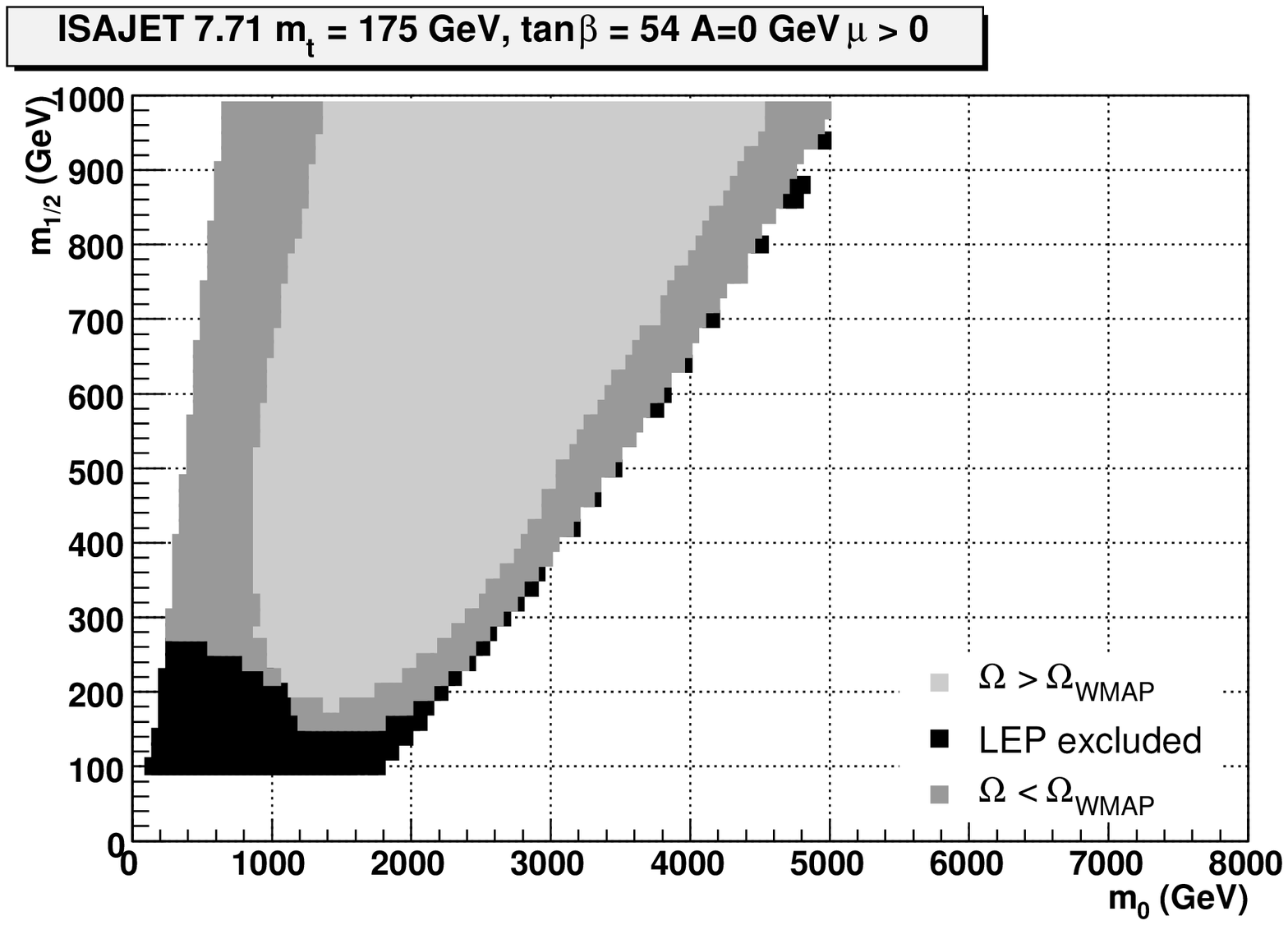}
\end{center} 
\caption{\label{scan4} Same as Fig.~\ref{scan1}, but for $\tan \beta = 50$  
(left plot) and $\tan \beta = 54$ (right plot).
}
\end{figure}

No focus point solution is found for $\tan \beta < 7$. For 
large values of $\tan \beta$, the focus point solutions move to 
lower values of $m_0$ as shown in Fig.~\ref{scan4}. For a fixed 
value of $m_{1/2}$, the gluino and gaugino masses are hardly affected. 
However, the mass of the scalars becomes smaller, and the production of 
$\tilde q \tilde g$ and  $\tilde q \tilde q$ pairs, followed by the 
squark cascade decay into gluinos, chargino and neutralinos, may be 
observed at LHC. Note also that the width of the coannihilation 
strip is heavily affected by the value of $\tan \beta$.

\begin{figure}[!p]
\begin{center}
\includegraphics[width=7cm]{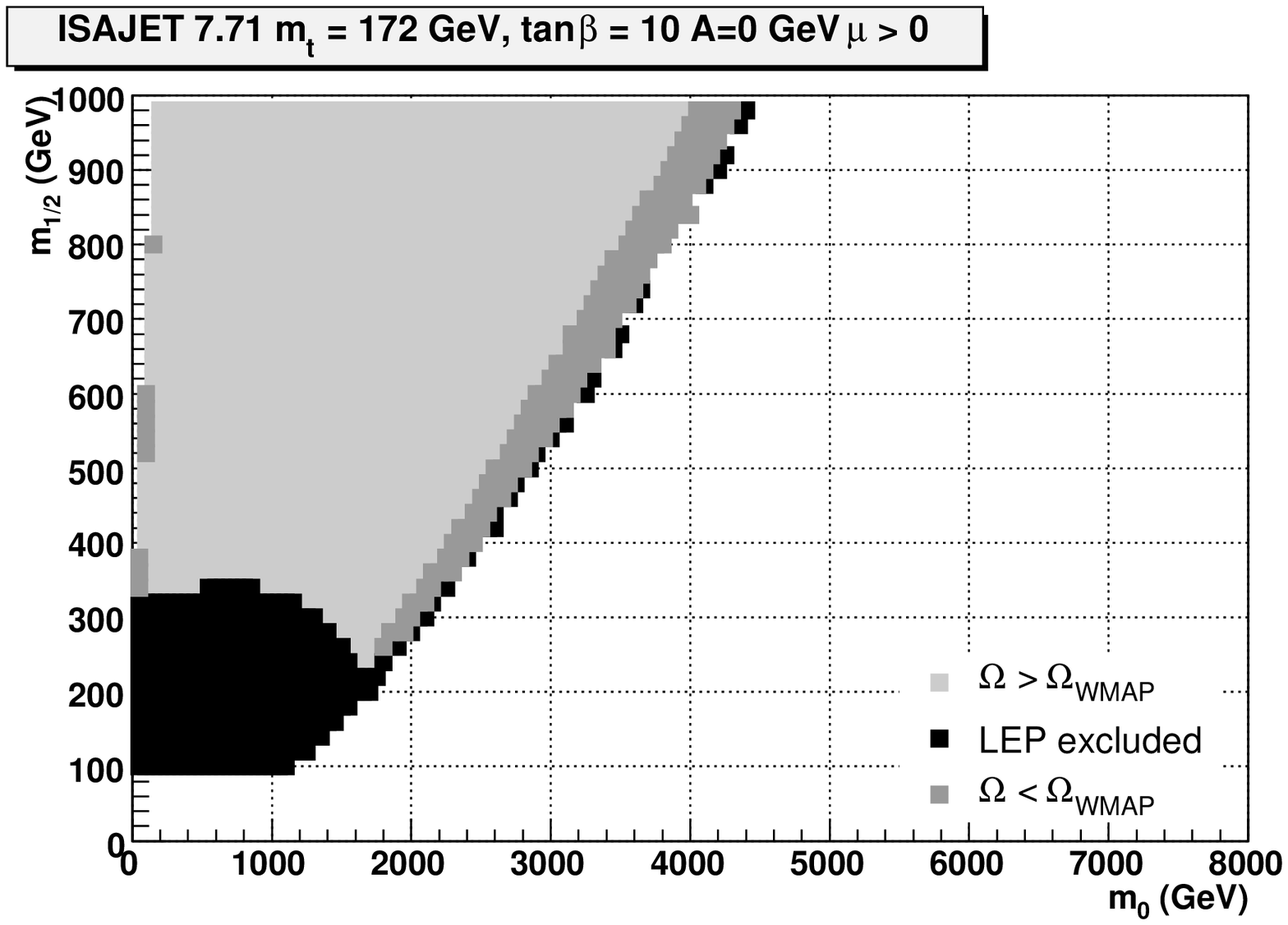}
\includegraphics[width=7cm]{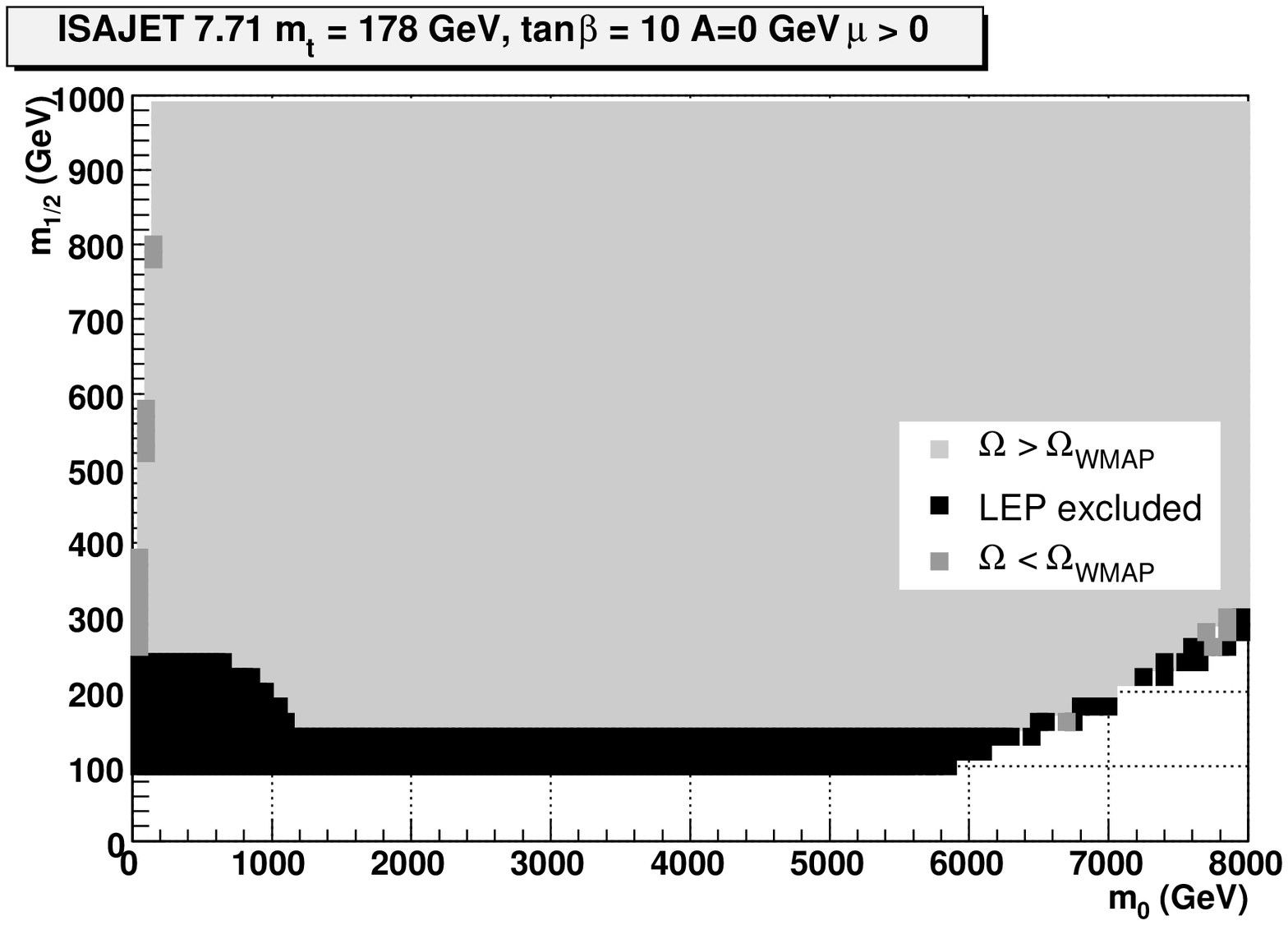}
\end{center} 
\caption{\label{scan5} Same as Fig.~\ref{scan1}, but for a top mass of 
172~GeV (left plot) and 178~GeV (right plot).
}
\end{figure}

\begin{figure}[!p]
\begin{center}
\includegraphics[width=9cm]{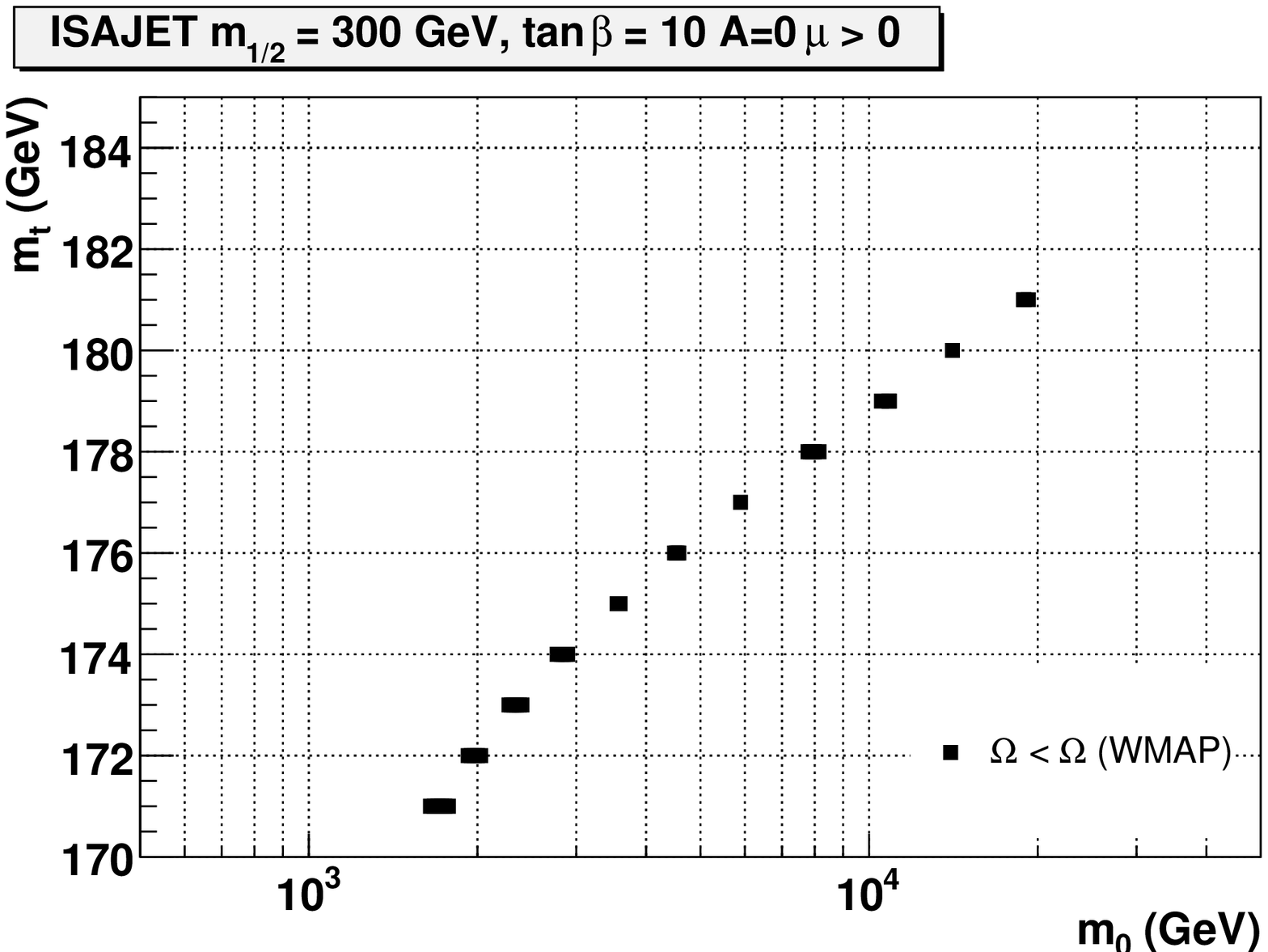}
\end{center} 
\caption{\label{scan6} The picture shows the values of the 
mSUGRA common scalar mass $m_0$ and the top mass $m_t$  
which have a neutralino relic density compatible with cosmological 
measurements. The values of 
$m_{1/2}=300$~GeV, $\tan \beta = 10$, $A_0 = 0$, and a positive $\mu$
were used. The RGE were solved using ISAJET. 
}
\end{figure}

The position of the focus point strip is also very 
sensitive to the value of the 
top mass. A lighter top pushes the focus point strip to smaller values of 
$m_0$ (Fig.~\ref{scan5}). The relation between the value of the scalar mass 
$m_0$ for which a focus point solution is found and the top mass is 
shown in Fig.~\ref{scan6} for fixed values of $\tan \beta = 10$, $A = 0$, 
$\mu > 0$ and $m_{1/2} = 300$~GeV.

The location of the focus point strip in the $(m_0,m_{1/2})$ is much less 
sensitive to the choice of the sign of $\mu$ or the value of $A$, at least 
in the range between -1000~GeV and 1000~GeV.

\begin{table}[!tbp]
\begin{center}
 \begin{tabular}{||c|c|c|c||} 
\hline 
Particle  & Mass (GeV) & Particle & Mass (GeV) \\
\hline 
$\tilde \chi^0_1$    & 103.35 & $\tilde e_L$ & 3547.5 \\
$\tilde \chi^0_2$    & 160.37 & $\tilde e_R$     & 3547.5 \\
$\tilde \chi^0_3$    & 179.76 & $\tilde \nu_e$   & 3546.3 \\
$\tilde \chi^0_4$    & 294.90 & $\tilde \tau_1$  & 3519.6 \\
$\tilde \chi^{\pm}_1$ & 149.42 & $\tilde \tau_2$  & 3533.7 \\
$\tilde \chi^{\pm}_2$ & 286.81 & $\tilde \nu_{\tau}$ & 3532.3 \\
{\em \~ g}    & 856.59 & $h$           & 119.01 \\
$\tilde u_L$ & 3563.2 &$H^0$         & 3529.7 \\
$\tilde u_R$ & 3574.2 &$A^0$         & 3506.6 \\
$\tilde b_1$ & 2924.8 &$H^{\pm}$     & 3530.6 \\
$\tilde b_2$ & 3500.6 & & \\
$\tilde t_1$ & 2131.1 & & \\
$\tilde t_2$ & 2935.4 & & \\
\hline
\end{tabular}
\end{center}
\caption{\label{tab1} The mass spectrum of the benchmark 
focus point described in the text.}
\end{table}

From the considerations above the 
following point in the parameter space was chosen for the detailed 
study reported in the next sections:

\begin{displaymath}
m_0 = 3550 \; \mbox{GeV}, m_{1/2} = 300 \; \mbox{GeV}, A = 0 \;\mbox{GeV}, 
\mu > 0, \tan \beta = 10
\end{displaymath}

with the top mass set to 175~GeV and the mass spectrum computed with 
ISAJET. In Table~\ref{tab1} the mass spectrum for this 
point is given. The scalar partners of Standard Model fermions
have a mass between 2131~GeV and 3574~GeV. The neutralinos and charginos 
have masses between 103.3~GeV and 294.9~GeV. The gluino is the lightest 
strongly interacting state, with a mass of 856.6~GeV. The lightest Higgs boson 
has a mass of 119~GeV, while the other Higgs states have a mass 
at more than 3~TeV, well beyond the expected LHC reach. 

At this point of the mSUGRA parameter space, the total SUSY production 
cross section at the LHC, as computed by HERWIG~\cite{HERWIG} at leading 
order, is 5.00~pb. It is dominated by the production 
of gaugino pairs, $\tilde \chi^0 \tilde \chi^0$ (0.22~pb), 
$\tilde \chi^0 \tilde \chi^{\pm}$ (3.06~pb),
and $\tilde \chi^{\pm} \tilde \chi^{\pm}$ (1.14~pb). 

The production of gluino pairs (0.58~pb) is also significant. 
The gluino decays into $\tilde \chi^0 q \bar q$ (29.3\%), 
$\tilde \chi^0 g$ (6.4\%), or $\tilde \chi^{\pm} q \bar q^{'}$ (54.3\%).
The quarks in the final state belong to the third generation 
in 75.6\% of the decays. 

The more recent Tevatron data favor a lighter top mass~\cite{TevTop}.
As the location of the focus point region in parameter space is quite 
sensitive to the value of the top mass, it is opportune to discuss 
how the phenomenology of our benchmark would change by using a 
lower value of 172 GeV for the top mass. The correct relic density can 
be obtained lowering the value of $m_0$ to 2000 GeV. The gluino would then 
have a slightly lower mass of 818~GeV. 
The neutralino and chargino 
masses are also slightly smaller, but remain within 10~GeV of the 
values of those of our benchmark point. The leptonic decays of the 
$\tilde \chi^0_2$ and $\tilde \chi^0_3$, which are the basis of the 
analysis discussed in section~\ref{sec4}, still occur but with a smaller 
branching ratios (1.9\% and 3.3\% instead of 3.3\% and 3.8\%). 

In this scenario, the event rates would be slightly different than in our 
benchmark scenario, but the procedures and the general conclusions of 
the studies presented 
in the next sections would still hold.  

In addition, the lower value of $m_0$ would probably open up the 
possibility to discover the scalar quarks at the LHC . 

\section{Inclusive searches}
\label{sec3} 

In this section, the possibility to detect an excess over Standard Model 
expectations in the production of events with hard jets and large missing 
energy is investigated. The cuts 
will be optimized to be sensitive to the production of gluino pairs, 
but they will not rely on any specific decay of gluino or its daughters.

The production of Supersymmetry events at the LHC was simulated using 
HERWIG 6.55~\cite{HERWIG}. The top background was produced using 
MC@NLO~2.31~\cite{MCNLO}. The fully inclusive $t\bar t$ production was 
simulated. This is expected to be the dominant Standard Model background 
for the analysis presented here.
The Z+jets and the W+jets backgrounds were produced with 
ALPGEN 2.05~\cite{ALPGEN} for the hard process, requiring at least 
2 jets with $p_T>40$~GeV and $|\eta| < 6$,
and HERWIG for subsequent 
parton shower and hadronization. The vector bosons were 
forced to decay into leptons. The $b\bar b$+jets background was simulated 
using ALPGEN for the hard process, requiring $|\eta| <5$ and 
$p_T>40$~GeV for the $b$ quarks and $|\eta| < 3$ and $p_T>40$~GeV
for the additional jets (at least one). HERWIG was used for the simulation
of the subsequent parton shower and hadronization.
The QCD light jet background is expected to be negligible for the analysis 
presented here, which require either two tagged $b$-jets or isolated leptons.
In order to verify this assumption, the production of 
events with 3 or more jets with $p_T>40$~GeV was simulated using ALPGEN 
for the hard process and HERWIG for parton shower and 
hadronization. All jet flavours except $b$-jets were included.

The events were then processed by ATLFAST~\cite{ATLFAST} to simulate 
the detector response. 

\begin{table}[!tbp]
\begin{center}
\begin{tabular}{||c|c|c|c|c||} 
\hline 
Process  & Generator & Events ($10^6$) & $\int L \mbox{d} t (\mbox{fb}^{-1})$ 
         & $L (\mbox{cm}^{-2} \mbox{s}^-1$) \\
\hline 
SUSY                & HERWIG & 0.15 &  30  & $10^{33}$ \\
SUSY                & HERWIG & 1.5  & 300  & $10^{34}$ \\
$t \bar t$          & MC@NLO & 16.7 &  22  & $10^{33}$ \\
$W$+jets            & ALPGEN & 18.2 &  18  & $10^{33}$ \\
$Z$+jets            & ALPGEN &  6.3 &  20  & $10^{33}$ \\
$bbq$               & ALPGEN & 10.9 &  0.49& $10^{33}$ \\
$bb+N$ jets ($N>1$) & ALPGEN & 11.6 &  2.4 & $10^{33}$ \\
$N$ jets ($N>2$)    & ALPGEN &  2.7 & 0.0016 & $10^{33}$ \\
\hline
\end{tabular}
\end{center}
\caption{\label{tab2} Simulated data samples used for the fast simulation 
studies reported here.}
\end{table}

A list of the simulated data used for fast simulation studies is 
shown in Table~\ref{tab2}. The different instantaneous luminosities 
are taken into account through the parametrization used by ATLFAST 
to simulate the detector response, 
since during operation at low luminosity ($10^{33} \; \mbox{cm}^{-2} 
\mbox{s}^{-1}$) the detector performance is not degraded by pile-up effects.
The standard parametrization of the ATLAS $b$-tagging performance was used, 
which assumed a $b$-tagging efficiency of 0.6 at low luminosity and 
0.5 at high luminosity, for a rejection factor (the inverse of efficiency) 
for charm and light quark jets of 10 and 100 respectively.
Muons and electrons are considered isolated if they are separated by 
the closest 
calorimeter cluster in the plane of pseudorapidity and azimuthal angle 
by at least $\Delta R = \sqrt{\Delta \eta^2 + \Delta \phi^2} > 0.4$ and if 
the transverse energy measured by the calorimeter in a cone 
centered on the lepton and of width $\Delta R = 0.2$ 
is not larger than 10 GeV (excluding the energy of the lepton itself).

Trigger efficiencies were not taken into account by the detector simulation. 
The cuts of the analyses described in this and in the following 
sections are more stringent than the selections of the ATLAS trigger 
menu foreseen for operation at the design luminosity of LHC~\cite{trigTDR}.
In particular, the cuts on jets and missing energy are more severe than the 
trigger requirement of at least one jet with $p_t > 70$~GeV and 
$E^T_{Miss} > 70$~GeV. Many of the events used in the analysis
presented in the next section, with two isolated leptons in the final state, 
would also be selected by the electron and muon triggers.
A study of the impact of trigger efficiencies on event rates, as well as 
the inclusions of detector effects not described by the 
parametrized simulation is outside the scope of this document. 

The analysis presented in this paper all require a transverse 
missing energy of at least 70~GeV or more, and either two jets 
with $p_T > 30$~GeV tagged as $b$-jets or isolated leptons. 
With these cuts, the main contribution from the QCD multi-jet 
background is expected to come from the $b \bar b$+jets 
events. This is demonstrated by the results reported in Table~\ref{tabQCD}, 
which show that after these preliminary cuts mentioned above, the light 
jets production is reduced to less than 10\% of the heavy flavour background.  
In the following, the light jet background would not be considered.

\begin{table}[!h]
\begin{center}
\begin{tabular}{||c|c|c|c|c||} 
\hline 
Process    & Events      & $E^T_{\mbox{Miss}}$ cut & 2 $b$ jets & 1 lepton \\ 
\hline
bb+jets    &     94581.3 & 344.296 & 106.074 & 20.8 \\
light jets & 2.69346e+06 & 193.282 &  4.8    & 1.6  \\
\hline
\end{tabular}
\end{center}
\caption{\label{tabQCD} Contributions of the multi jet background to the 
signatures studied in this paper,
evaluated with ATLFAST events for low luminosity operation. The number of 
events corresponds to an integrated luminosity of 1.6~$\mbox{pb}^{-1}$.
The third column shows the number of events which have 
$E^T_{\mbox{Miss}} > 70$~GeV. The fourth column reports the 
number of events after the additional requirement of 2 tagged $b$ jets, 
the last column the number of events with $E^T_{\mbox{Miss}} > 70$~GeV
and one isolated electron or muon.}
\end{table}

The most abundant gluino decay modes are 
$\tilde g \rightarrow \tilde \chi^0 t \bar t$ (27.9\%), 
$\tilde g \rightarrow \tilde \chi^{+} t \bar b$ (22.0\%) and 
$\tilde g \rightarrow \tilde \chi^{-} \bar t b$ 
(22.0\%)\footnote{In the following, whenever the decay 
$\tilde g \rightarrow t \bar b \tilde \chi^-$ will be mentioned, the 
charge conjugate is implied.}.
Events with gluino pair production 
have thus at least four hard jets, and may have many more additional jets 
because of the top hadronic decay modes and the chargino and neutralino 
hadronic decay modes, such as $\tilde \chi^0_2 \rightarrow \tilde 
\chi^0_1 q \bar q$ or $\tilde \chi^{\pm}_1 \rightarrow \tilde 
\chi^0_1 q \bar q^{'}$
. When both gluinos decay to third generation quarks,which happens 
for 57.2\% of the events, at least 4 jets are
$b$-jets. A missing energy signature is provided by the two 
$\tilde \chi^0_1$ in the final state, and possibly by neutrinos coming 
from the top quark and the gaugino leptonic decay modes.  

These events can be separated from the Standard Model background 
requiring the presence of 
hard jets and large missing transverse energy. The request of 
$b$-jets in the final state suppresses the W+jets and Z+jets background 
so that the dominant surviving Standard Model background is 
the top pair production and the $bb$+jets production. This request 
also enhances the signal with respect to these backgrounds, as most 
signal events have four true b-jets (rather than two) in the final 
state.

The following selections are made:

\begin{itemize}

\item At least one jet with $p_T > 120$~GeV

\item At least four jets with $p_T > 50$~GeV, and at least two of them 
tagged as $b$-jets.

\item $E_{MISS}^T > 100$~GeV

\item $E_{MISS}^T/M_{EFF} > 0.12$

\end{itemize}

Here, the effective mass $M_{EFF}$ is defined as the scalar sum of the 
transverse missing energy and the transverse momentum of all 
the reconstructed hadronic jets. The fraction $s = E_{MISS}^T/M_{EFF}$ 
measures the relative importance of the missing energy and the 
hadronic jet components of $M_{EFF}$. In
events due to the production of a gluino pair this ratio is on the average 
larger than in the background processes. 

\begin{table}[!tbp]
\begin{center}
\begin{tabular}{||c|c|c|c|c||} 
\hline 
Sample           & Events & Basic cuts & 2 $b$-jets & $M_{eff} > 1600$ GeV\\
\hline 
SUSY ($\tilde \chi\tilde \chi$) &   44200           &    359 &    20 &   0 \\
SUSY ($\tilde g \tilde g$)      &    5800           &   3522 &  1625 & 508 \\
$t \bar t$                      & $ 7.6 \cdot 10^6$ & 174993 & 39816 & 506 \\
$W$+jets                        & $10.1 \cdot 10^6$ &  62546 &   397 &  18 \\
$Z$+jets                        & $3.15 \cdot 10^6$ &  45061 &   306 &  20 \\
$bb$+jets                       & $272  \cdot 10^6$ &  39579 & 12124 & 141 \\
\hline
\end{tabular}
\end{center}
\caption{\label{tab3} Efficiency of the cuts used for the inclusive search, 
evaluated with ATLFAST events for low luminosity operation. The number of 
events corresponds to an integrated luminosity of 10~$\mbox{fb}^{-1}$.
The third column shows the number of events which pass the cuts on jet 
transverse energy, transverse missing energy, and $E_{MISS}^T/M_{EFF}$ 
reported in the text. The number of events shown in 
the fourth column is obtained after the additional 
requirement that two of the 
jets are tagged as $b$-jets. 
Finally, the number of events which passes all the selection and have an 
effective mass larger than 1600~GeV is shown in the last column.}
\end{table}

The efficiency of these cuts is reported in Tab.~\ref{tab3}. The SUSY events 
are divided in gluino pair and gaugino pair production. The 
selection efficiency for the $\tilde g \tilde g$ events is 
60.7\% after all the cuts except the request of two $b$-tagged jets. 
This request reduces the selection efficiency to 28.0\%. 
The ratio between the number of $\tilde \chi\tilde \chi$ 
and $\tilde g\tilde g$ events is already suppressed by 
two orders of magnitude by the selections on hard jets and 
missing energy and it becomes negligible after the request of 2 $b-$jets.

The requirement of two $b$-jets dramatically reduces the 
$W$+jets and $Z$+jets backgrounds. Because of the large number of true 
$b$ jets in the SUSY events, it also reduces the signal less than the 
$t \bar t$ and $b \bar b$ backgrounds. These are the dominant sources 
of Standard Model backgrounds after all selections.

\begin{figure}[!htbp]
\begin{center}
\includegraphics[width=10cm]{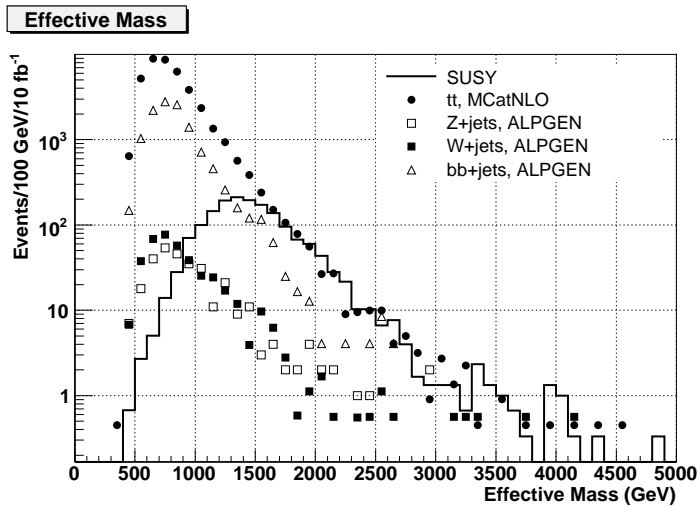}
\end{center} 
\caption{\label{inclusive} Distribution of the effective mass defined in 
the text, for SUSY events and the Standard Model background, for  
an integrated luminosity of 10~$\mbox{fb}^{-1}$.}
\end{figure}

The distribution of the effective mass after these selection cuts is 
reported in Fig.~\ref{inclusive}. The statistics corresponds to an 
integrated luminosity of 10~$\mbox{fb}^{-1}$. The number of signal and 
background events 
with an effective mass larger than 1600~GeV is comparable and is 
reported in the last column of Table~\ref{tab3}. 

From these numbers, it is possible for the focus point scenario 
under study to estimate the minimum integrated 
luminosity required to observe a deviation from Standard Model expectations.
The statistical significance of the signal from Supersymmetry
is $S/\sqrt{B} = 6.1$ for an integrated luminosity of 1~$\mbox{fb}^{-1}$. 
These numbers assume nominal detector performances (that is, detector 
commissioning has been completed) and includes only statistical 
errors. In order to reach this sensitivity the systematic error on the 
background, whose rate is similar to that of the signal after all cuts and 
at large effective mass, will need to be reduced to a level well below 
$\sigma_{\mbox{stat}} (B)/B = \sqrt{B}/B = 12\%$ (for 1~$\mbox{fb}^{-1}$ of 
integrated luminosity). 

\section{The di-lepton edge}
\label{sec4}

As discussed at the end of Section~\ref{sec2}, at the 
selected benchmark point the neutralinos are produced
either directly or by gluino decays. Despite the lower cross section, 
the latter mechanism dominates after the cuts on missing energy and 
jets, which are necessary to remove the Standard Model backgrounds.  
The leptonic decays of the second and third neutralino

\begin{equation}
\label{eq1}
\tilde \chi^0_2 \rightarrow \tilde \chi^0_1 l^+ l^-
\end{equation}

\begin{equation}
\label{eq2}
\tilde \chi^0_3 \rightarrow \tilde \chi^0_1 l^+ l^-
\end{equation}

occur with a branching ratio of 3.3\% and 3.8\% per lepton flavour 
respectively. The two leptons in the final state provide a  
clear signature. Their invariant mass distribution has a 
kinematic endpoint value equal to the mass difference of the two neutralinos 
involved in the decay, which is 

\begin{equation}
\label{eq3}
m_{\tilde \chi^0_2} - m_{\tilde \chi^0_1} = 57.02 \; \mbox{GeV} \;\; 
m_{\tilde \chi^0_3} - m_{\tilde \chi^0_1} = 76.41 \; \mbox{GeV}
\end{equation}

\begin{figure}[!htbp]
\begin{center}
\includegraphics[width=10cm]{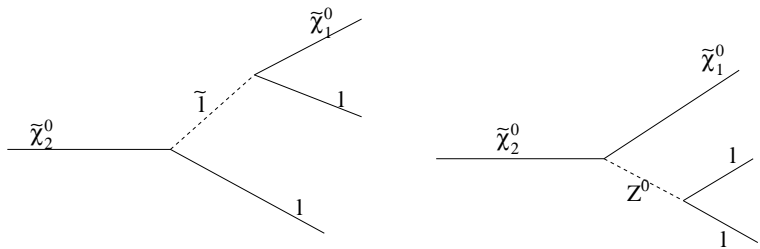}
\end{center} 
\caption{\label{Fey} Feynman diagrams of the $\tilde \chi^0_2$ leptonic 
decay. The decay of the $\tilde \chi^0_3$ proceeds according to a similar
diagram.}
\end{figure}

The fourth neutralino is heavy enough that the decay 
$\tilde \chi^0_4 \rightarrow \tilde \chi^0_1 Z^0$ is open; the leptons 
produced by this decay have an invariant mass equal to the $Z$ mass and 
do not allow to measure the neutralino masses. 

The leptonic decays of $\chi^0_2$ and $\chi^0_3$ 
proceed according to the two diagrams  
reported in Fig.~\ref{Fey}. The resulting distribution of the 
lepton four momenta is given in Ref.~\cite{Mih99}. In the focus 
point region, the diagram with the 
virtual slepton exchange is negligible, because of the large slepton 
mass. Assuming that only the diagram with the $Z$ exchange contributes, 
and neglecting the mass of the leptons in the final state, 
we get the following expression~\cite{Mon05} for the distribution of the 
two lepton invariant mass:

\begin{equation}
\label{eq4}
\frac{d\Gamma}{dm} = C m \frac{\sqrt{m^4 -m^2(\mu^2+M^2)+(\mu M)^2}}
{(m^2-m^2_Z)^2} [ -2m^4+m^2(2M^2+\mu^2)+(\mu M)^2 ]
\end{equation}

In this formula, $C$ is a normalization constant, 
$\mu=m_2-m_1$ and $M=m_2+m_1$, where $m_1$ and $m_2$ are the 
signed mass eigenvalues of the daughter and parent neutralino respectively. 
At the focus point, the mass eigenvalues  of the two lightest neutralinos 
have the same sign, while the $\tilde \chi_0^3$  has the opposite sign. 
In the decay of the $\tilde \chi_0^2$ it is thus 
$\mu = m(\tilde \chi^0_2) - m(\tilde \chi^0_1)$ and 
$M = m(\tilde \chi^0_2) + m(\tilde \chi^0_1)$. 
In the decay of the $\tilde \chi_0^3$ the role of $\mu$ and $M$ is 
inverted~\footnote{The second term in parenthesis in Eq.~\ref{eq4}, 
$m^2(2M^2+\mu^2)$,  
is not invariant for the exchange of $\mu$ and M. This term 
comes from an interference term in the matrix element of the decay 
and does depend on the relative sign of the two neutralino eigenstates.}.

\begin{figure}[!p]
\begin{center}
\includegraphics[width=7cm]{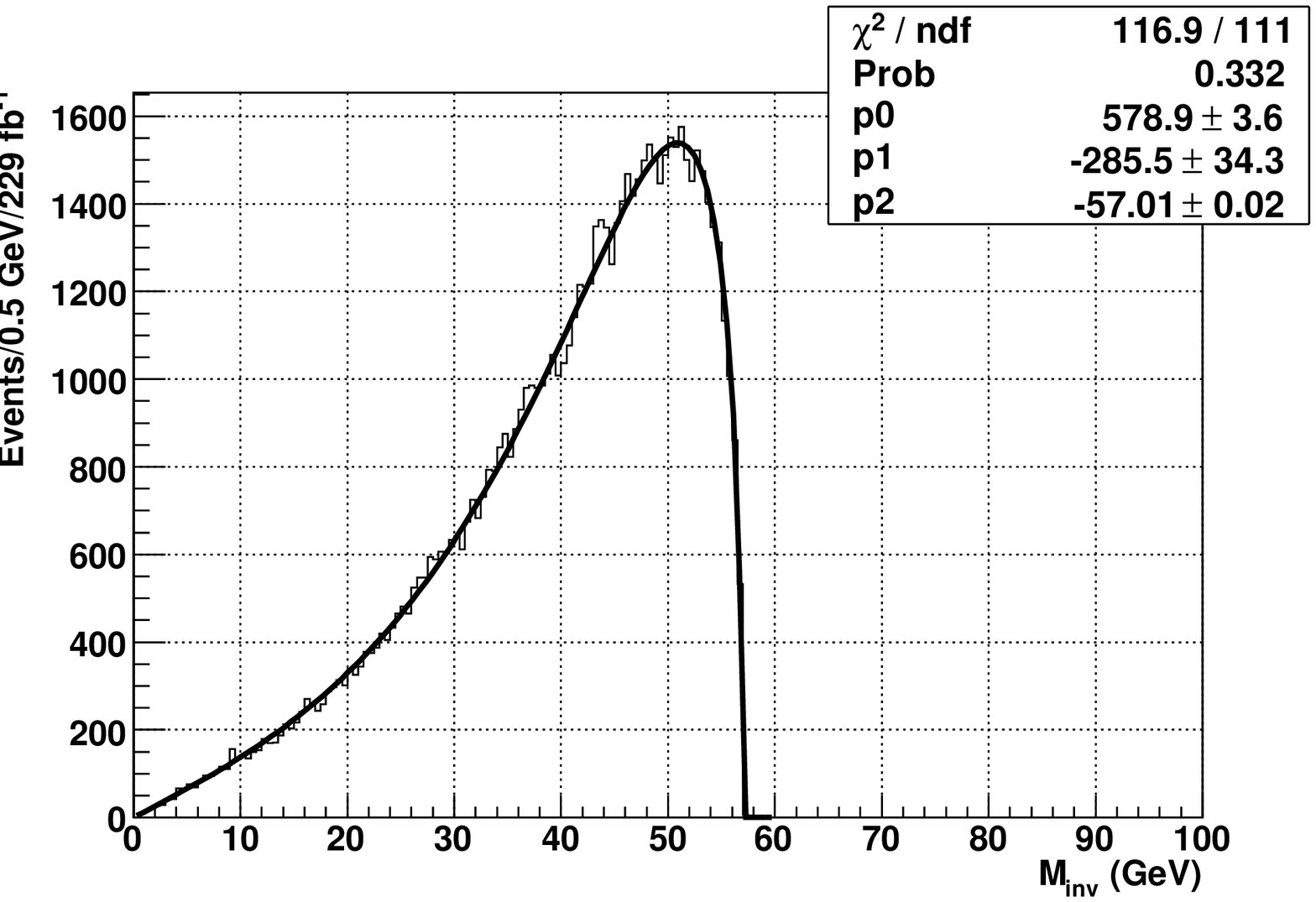}
\includegraphics[width=7cm]{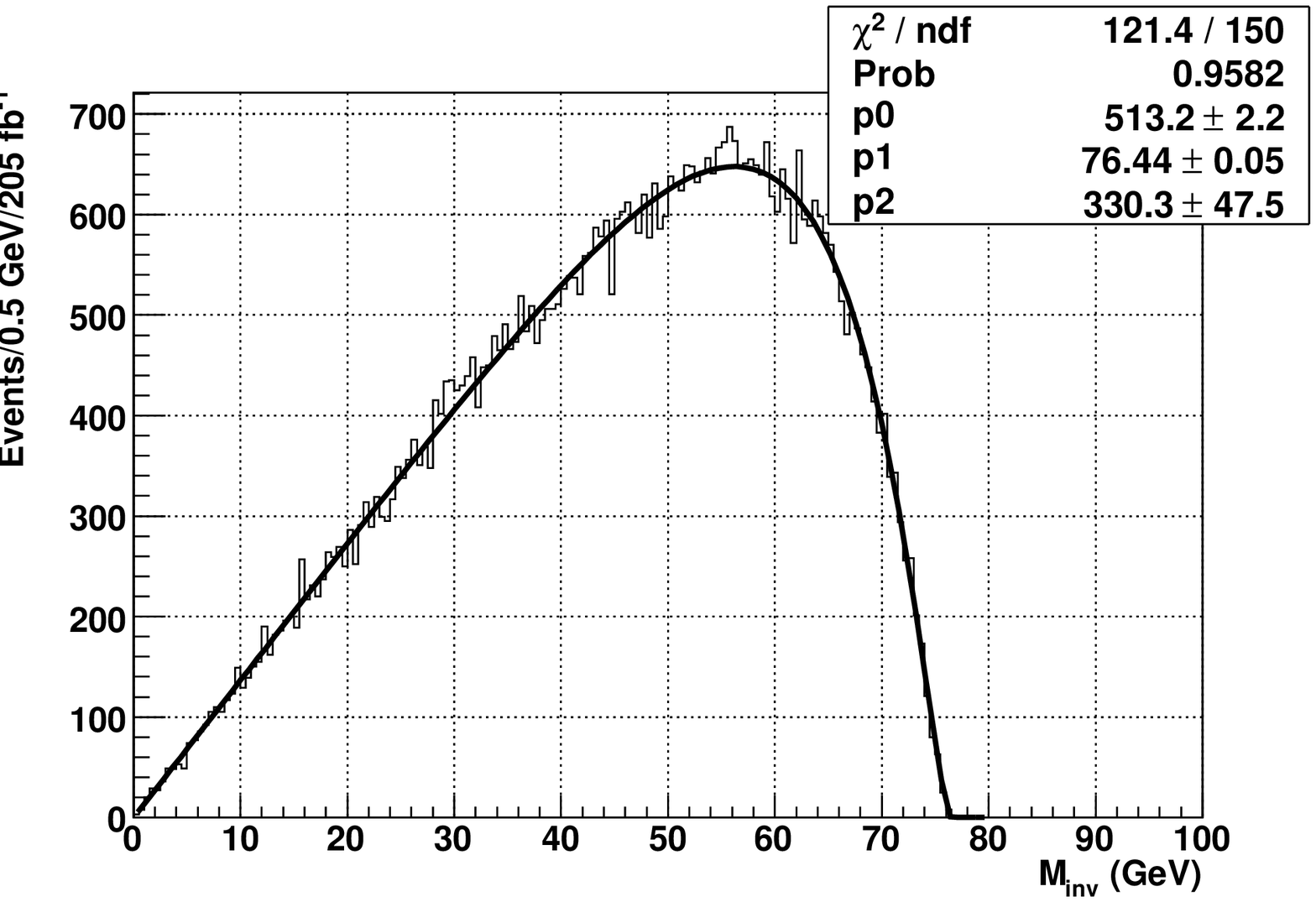}
\end{center} 
\caption{\label{lep0} Distribution of the invariant mass of lepton pairs 
from the $\tilde \chi^0_2 \rightarrow \tilde \chi^0_1 l l$ decay (left plot) 
and the $\tilde \chi^0_3 \rightarrow \tilde \chi^0_1 l l$ decay (right plot).  
The histogram is the HERWIG Monte Carlo distribution, the line is a
fit performed with the analytical formula described in the text.}
\end{figure}

The shape of the distribution is different in the two cases, as can 
be seen from Fig.~\ref{lep0} where the distribution of the invariant 
mass of the two leptons from the two decays is shown. The histogram 
is the true invariant mass of the two leptons and the line is a fit performed 
with the function of eq.~\ref{eq4}. 

The fit provides both the difference and the sum of the 
neutralino masses. The latter is however much less precise, since the 
dependence of the distribution shape on the lightest neutralino mass
becomes very weak when this mass is larger than the difference between the 
masses of the neutralinos.
The results of the fit to the Monte Carlo distribution are in agreement with 
the true values of the neutralino masses, 
with the exception of the $m(\tilde \chi^0_2) - m(\tilde \chi^0_1)$ 
mass difference, which 
is precise enough to be affected by the mass of the muon, which is neglected 
in eq.~\ref{eq4}. 
The formula neglects the distortion of the shape caused by the 
kinematical cuts. The systematic errors on the mass difference arising 
from      the analysis cuts on the lepton 
transverse momentum and pseudorapidity 
have been estimated as follows. 
When the kinematical cuts of the analysis on the lepton momenta 
($p_T > 10$~GeV and $|\eta| < 2.5$) are imposed, the 
value of the endpoints obtained from the fit of the 
Monte Carlo Truth distribution changes by 0.20 GeV for the 
$m(\tilde \chi^0_2) - m(\tilde \chi^0_1)$ 
difference and by 0.04~GeV for the $m(\tilde \chi^0_3) - m(\tilde \chi^0_1)$
difference. This was taken as the systematic error induced on the 
endpoint by the kinematical cuts.

The analysis of the simulated data was performed with the following 
selections:

\begin{itemize}

\item Two isolated leptons with opposite charge and same flavour with 
$p_T > 10$~GeV and $|\eta| < 2.5$

\item $E^T_{MISS} > 80$~GeV, $M_{EFF} > 1200$~GeV, 
$E^T_{MISS}/M_{EFF} > 0.06$

\item At least one jet with $p_T > 80$~GeV, at least four jets 
with $p_T > 60$~GeV, and at least six jets with $p_T > 40$~GeV

\end{itemize}

\begin{figure}[!p]
\begin{center}
\includegraphics[width=10cm]{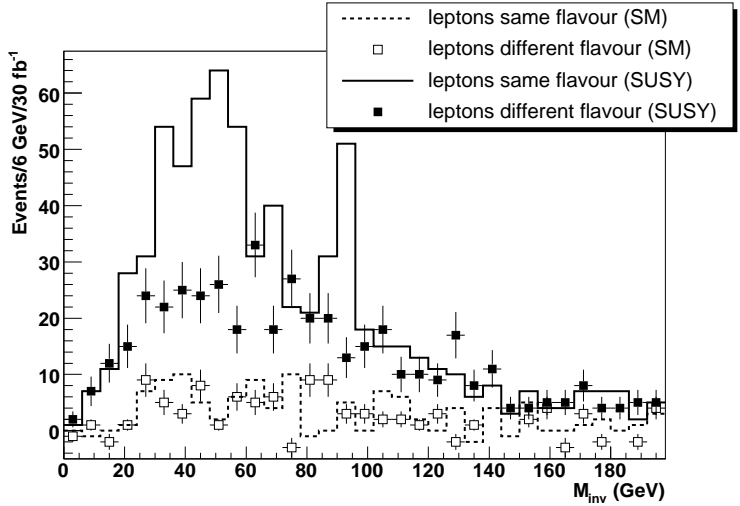}
\end{center} 
\caption{\label{lep1} The full and dashed lines are the 
distribution of the invariant mass of lepton pairs with the same 
flavour and opposite charge for SUSY events and the Standard Model 
background respectively.
The full markers (Supersymmetry) and the empty markers (Standard Model)
are the distribution of invariant mass of the lepton pairs with opposite 
flavour and opposite charge. The number of events correspond to 
an integrated luminosity of 30~$\mbox{fb}^{-1}$.}
\end{figure}

The efficiency of the various cuts is shown in Table~\ref{tab4} 
for low-luminosity running conditions and an integrated statistics 
of 30~$\mbox{fb}^{-1}$. After all cuts, 411 SUSY and 83 Standard Model 
events are left with a 2-lepton invariant mass smaller than 80~GeV. 
In 247 of the SUSY events the decay (\ref{eq1}) or 
(\ref{eq2}) is indeed present in the Monte Carlo Truth record. 
In Fig.~\ref{lep1} the distribution of the invariant mass of lepton pairs 
after all cuts is shown for the signal and the Standard Model background 
(full and dashed lines respectively).

With these selections, the significance $SUSY/\sqrt{SM}$ would 
be 8.2 with an integrated luminosity of $1 \mbox{fb}^{-1}$ 
(neglecting systematic errors), which makes this channel competitive 
with the inclusive search in terms of discovery reach.

\begin{table}[!tbp]
\begin{center}
\begin{tabular}{||c|c|c|c|c||} 
\hline 
Sample    & Events & after cuts & $M_{ll} < 80$~GeV \\
\hline 
$\tilde g \tilde g$ signal      &    1027           &    259 & 247 \\ 
$\tilde g \tilde g$ background  &   16490           &    358 & 159 \\
$\tilde \chi\tilde \chi$        &  132483           &      7 &   5 \\
$t \bar t$                      & $22.7 \cdot 10^6$ &    131 &  77 \\
$W$+jets                        & $30.3 \cdot 10^6$ &      0 &   - \\
$Z$+jets                        & $9.45 \cdot 10^6$ &     18 &   6 \\
bb+jets                         & $817 \cdot 10^6$  &     12 &   0 \\
\hline
\end{tabular}
\end{center}
\caption{\label{tab4} Efficiency of the cuts used for the reconstruction 
of the neutralino leptonic decay, 
evaluated with ATLFAST events for low luminosity operation. The number of 
events corresponds to an integrated luminosity of 30~$\mbox{fb}^{-1}$.
The third column contains the number of events after all cuts, and the 
last column reports the number of events with a lepton invariant mass 
in the signal region.
SUSY events are divided in gluino pair production with the 
presence of either the $\tilde \chi^0_2 \rightarrow \tilde \chi^0_1 l^+ l^-$ 
or the $\tilde \chi^0_3 \rightarrow \tilde \chi^0_1 l^+ l^-$ decay (signal), 
gluino pair production
without these decays (background), and the $\tilde \chi \tilde \chi$ 
production.}
\end{table}

The background can be estimated from the data using the 
$e^+ \mu^-$ and $\mu^+ e^-$ pairs. In Fig.~\ref{lep1} the distribution of the 
lepton invariant mass is reported for the same flavour and the opposite 
flavour lepton pairs. Outside the signal region and the Z peak 
the two histograms are compatible. 
The opposite-flavour distribution was thus used to estimate 
and subtract the SUSY combinatorial 
and Standard Model backgrounds from the same flavour signal 
histogram.

\begin{figure}[!p]
\begin{center}
\includegraphics[width=10cm]{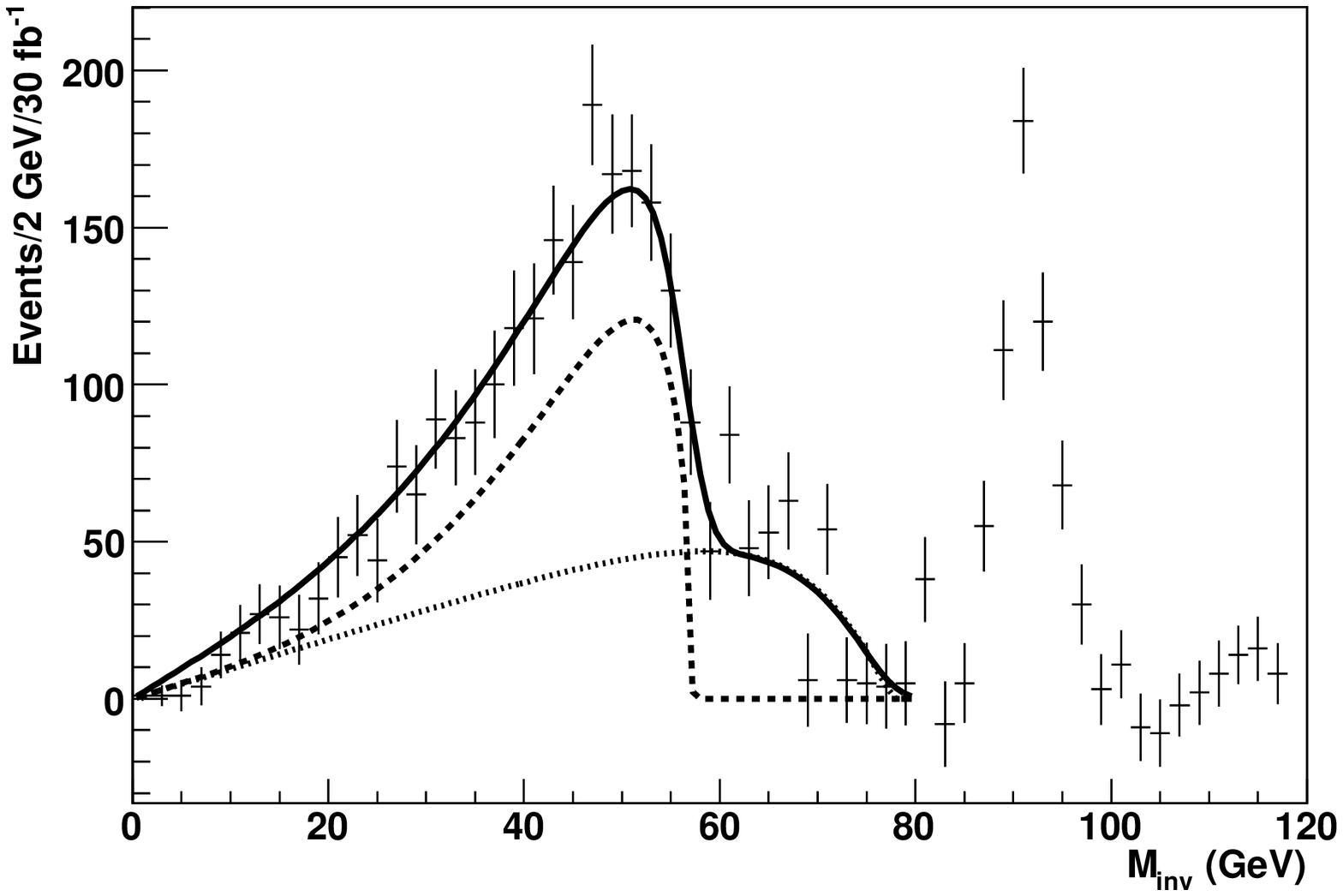}
\end{center} 
\caption{\label{lep4} Flavour-subtracted distribution of the 
invariant mass of lepton pairs, for 
an integrated luminosity of 300~$\mbox{fb}^{-1}$.
The fit function is superimposed as a full line; the contribution 
from the $\tilde \chi^0_2$ and $\tilde \chi^0_3$ decays are shown 
separately as a dashed and dotted line respectively. 
}
\end{figure}

We note that this technique may be used also in the search for a 
deviation from Standard Model prediction. Since part of the SUSY events 
is removed in the subtraction, the statistical significance would be 
smaller than that obtained with the number of events with opposite sign, 
same flavour leptons. However, the Standard Model contribution is canceled 
in the subtraction, removing many sources of systematics errors on 
background rates.

After the flavour subtraction and for a statistics of 30~$\mbox{fb}^{-1}$,
$179 \pm 32$ SUSY events and $19 \pm 37$ Standard Model events are 
left~\footnote{The quoted error is the uncertainty on the Monte Carlo 
prediction and it is not equal to the square root of the number of events.
The expected number of events is computed as
$b \times ( (N_{SF+} - N_{SF-}) -  (N_{OF+} - N_{OF-}))$ where $b$ is the 
factor needed to rescale the Monte Carlo statistics to 30~$\mbox{fb}^{-1}$, 
$N_{SF+}$ and $N_{SF-}$ are the number of same-flavour events with positive 
weight  and negative weight respectively, $N_{SF+}$ and $N_{SF-}$ are 
the corresponding numbers for opposite-flavour events. The error on 
this number is 
$b \times \sqrt{    N_{SF+} + N_{SF-}  +   N_{OF+} + N_{OF-} }$.
For the SUSY sample all events have positive weight and $b=1$ 
while for the top sample $b=1.36$ and events with both
positive and negative weights are present.}.
If the invariant mass of the lepton pair is required to be smaller than 
80~GeV, 
this leaves $170 \pm 25$ SUSY events and $8 \pm 13$ Standard Model events. 
The significance of the signal from neutralino leptonic decays is 2.5 for 
1~$\mbox{fb}^{-1}$ of integrated luminosity. 

After the flavour subtraction, the Standard Model background is compatible 
with 0, 
but still contributes to the invariant mass distribution by increasing the 
statistical fluctuations. The effect is small, since even before flavour 
subtraction the Standard Model contribution is smaller than the SUSY 
combinatorial 
background. For the high-luminosity studies reported in the rest of this 
section, 
the Standard Model contribution is not included. 

The flavour subtracted distribution is shown in Fig.~\ref{lep4}
for an integrated luminosity of 300~$\mbox{fb}^{-1}$ and high luminosity 
conditions. The presence of two edges is now apparent. 
The fit was performed with the sum of the $\tilde \chi^0_3$ and 
$\tilde \chi^0_2$ decay distributions provided by Eq.~\ref{eq4}, convoluted 
with a Gaussian smearing obtained from the width of the observed $Z$ peak. 
The fit parameters are the mass of the $\tilde \chi^0_1$ (which is the same 
for the two decays), the two mass differences 
$\tilde \chi^0_2 - \tilde \chi^0_1$ and 
$\tilde \chi^0_3 - \tilde \chi^0_1$, and the normalizations of the two decays.
A good fit $       \chi^2$ is only obtained using the correct 
values for the sign of the neutralino mass eigenstates. 

The values found for the two mass differences are 
$m(\tilde \chi^0_2) - m(\tilde \chi^0_1) = (57.2 \pm 0.4 \pm 0.2)$~GeV and 
$m(\tilde \chi^0_3) - m(\tilde \chi^0_1) = (78.1 \pm 1.4 \pm 0.04)$~GeV. 
The first error is the statistical one and the second is the systematic 
error due to the distortion of the distribution arising from the 
lepton kinematics cuts discussed earlier. They 
are compatible with the true values (eq.~\ref{eq3}). The mass of the 
lightest neutralino is not constrained by the fit, which gives
$m(\tilde \chi^0_1)= (0.3 \pm 2.1)$~TeV. 

The number of events due to each neutralino
decay can be computed as the integral of eq.~\ref{eq4}. The ratio of the 
events produced by the two decays can thus be determined from the 
parameters measured by the fit. Taking into account the correlation 
between the parameters, this gives $N_3/N_2 = 1.4 \pm 0.3$. 
This is compatible from the value $N_3/N_2 = 1.19$ which can be computed 
from the gluino, chargino and neutralino branching ratios.

\section{Reconstruction of the gluino decays}
\label{sec5}

The standard technique used to reconstruct the squark and gluino decays 
in mSUGRA is the combination of the leptons from the neutralino decays 
with the hardest jets in the event~\cite{ATDR,Gje02}. In models for which 
the two hardest jets in the event correspond to the 
quarks from the $\tilde q \rightarrow q \tilde \chi^0$ decay, it is possible 
to control the combinatorics from wrong jet associations and reconstruct 
all the masses of the particles in the $\tilde g \rightarrow q \tilde q 
\rightarrow \tilde \chi^0_2 q q \rightarrow \tilde l l qq \rightarrow 
\tilde \chi^0_1 ll qq$ decay chain. 

In the focus point region, however, the statistics of lepton pairs 
is not very high to begin with, and the   
three-body decays of the gluino and the presence of top quarks in most 
of these decays result in a large number of jets (between four and twelve, 
depending on the gluino and top decay modes) which comes from the 
decay of two gluinos. It is thus very difficult to control the 
resulting jet combinatorics in these events. 

\begin{figure}[!htbp]
\begin{center}
\includegraphics[width=7cm]{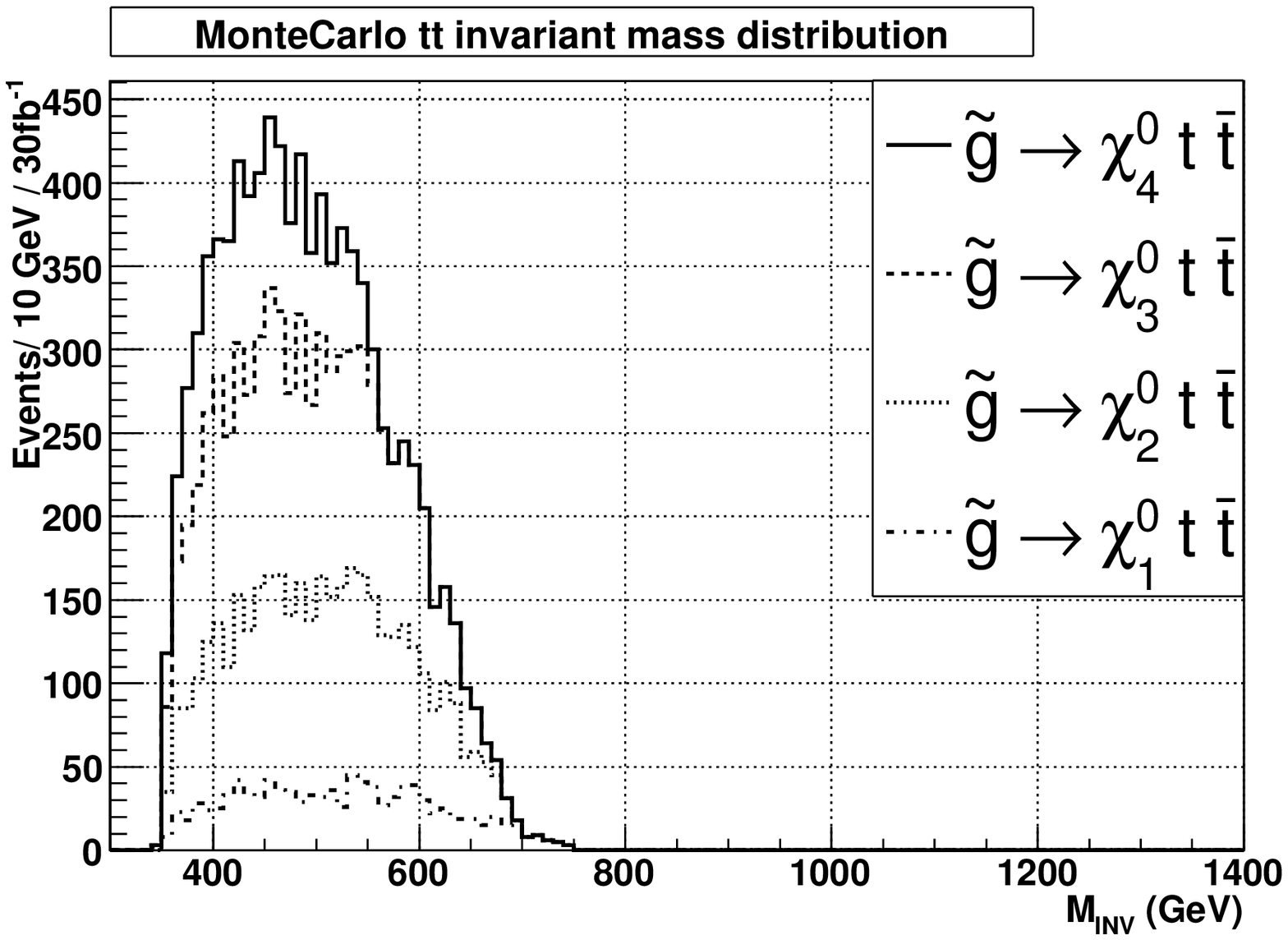}
\includegraphics[width=7cm]{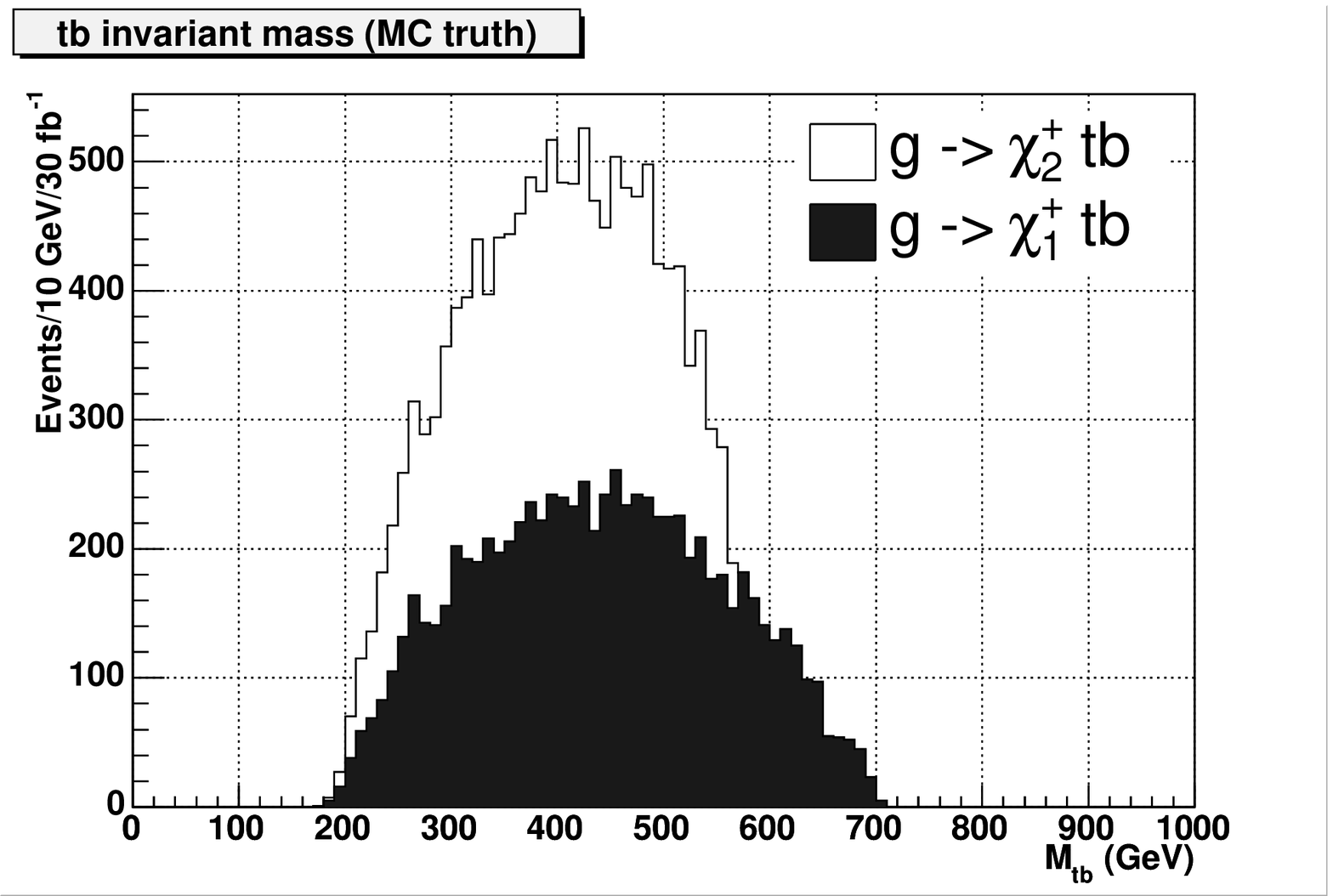}
\end{center}
\caption{\label{hq1} Parton-level invariant mass of the $t\bar t$ 
pairs from the 
$\tilde g \rightarrow tt \tilde \chi^0$ decays (left) and of the $t\bar b$ 
pairs from the $\tilde g \rightarrow t \bar b \tilde \chi^-$ decays (right). 
The contributions from the decay to the various neutralino
and chargino states are shown separately.}
\end{figure}

Instead, the strategy used was the explicit 
reconstruction of a top quark in the event using the hadronic decay mode, 
followed by the selection of $t \bar t$ and $t\bar b /\bar t b$ pairs, 
with appropriate kinematic
cuts (in particular, the angular separation) in order to 
reconstruct the $\tilde g \rightarrow t \bar t \tilde \chi^0$ and  
$\tilde g \rightarrow t \bar b \tilde \chi^-$ 
decays. 
The invariant mass distributions of these decays 
are shown at the parton level in Fig.~\ref{hq1}.
 
The distribution of the $t\bar t$ invariant mass has four end points, 
one for each neutralino state. Because of the poor experimental 
resolution on the top reconstructed momentum, they are unlikely to 
be separated from the data. The distribution falls almost linearly 
to the third endpoint at  $m(\tilde g) - m(\tilde \chi^0_2) = 696$~GeV, and 
only very few events due to the  $\tilde g \rightarrow \tilde \chi^0_1 
t \bar t$ decay are found at larger invariant mass.  

The distribution of the $t \bar b$ invariant mass has two endpoints at 
569.8~GeV and 707.2~GeV, corresponding to the difference between the 
mass of the gluino and that of the two charginos. 
Again, the first endpoint will be very difficult to extract 
from the data, because of the smearing from the finite 
jet energy resolution.

\subsection{Reconstruction of the $\tilde g \rightarrow \tilde 
\chi^0 t \bar t$ decay}

The reconstruction of the $t\bar t$ invariant mass using the decay 
$t\bar t \rightarrow jjb jj\bar b$ 
requires the presence of six jets, two of which 
tagged as $b$-jets to reduce the combinatorial background. 
Other jets are expected for signal events, from the decay of the other gluino. 
Thus, events were selected according to following cuts:

\begin{itemize}

\item $E^T_{MISS} > 120$~GeV

\item At least one jet with $p_T > 150$~GeV, at least 8 jets 
with $p_T > 40$~GeV, at least two of these tagged as 
$b$-jets.

\item $M_{EFF} >1200$~GeV


\end{itemize}

All the pairs of jets which were not $b$-tagged, with a transverse momentum 
$p_T > 30$~GeV, and with an invariant mass within $\pm 20$~GeV of the 
nominal W mass were used to build $W$ candidates. 
The combinatorial background is estimated from the events 
which contain jet pairs in the regions {\bf A:} 
$|m_{jj}-(m_W-30 \mbox{GeV})|<10$~GeV and {\bf B:} 
$|m_{jj}-(m_W+30 \mbox{GeV})|<10$~GeV. 
We call them the W ``sidebands''. The energy and momentum of 
the jet pairs of each sideband are then rescaled linearly 
by multiplying them by a factor 
$[m_W+2 (m_{jj}-(m_W \pm 30 \mbox{GeV}))]/m_{jj}$, 
so that their invariant mass lies in the W mass region
$m_W \pm 20 \mbox{GeV})$~\cite{His03}.

\begin{figure}[!htbp]
\begin{center}
\includegraphics[width=9cm]{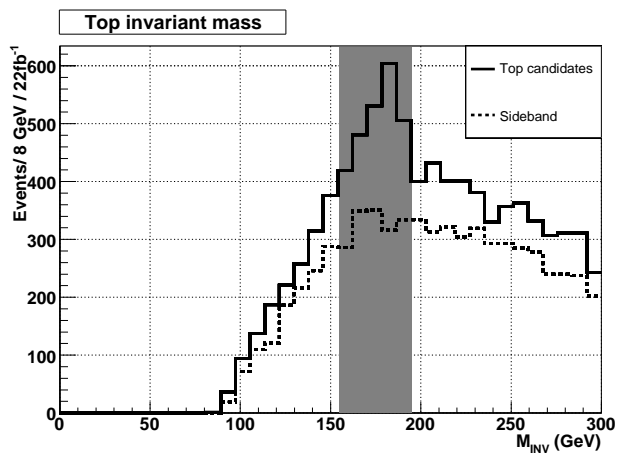}
\end{center} 
\caption{\label{topcandidates} Distribution of the invariant mass of the
top quark candidates (black line) and of the combinatorial background 
estimated from the data, as described in the text (dashed line). 
The plot corresponds to  22~$\mbox{fb}^{-1}$ of low-luminosity data and 
includes both the Supersymmetric signal and the Standard Model background.
The shaded area indicates the window used to select good top candidates.}
\end{figure}

Top candidates were found by combining each $W$ candidate with one $b$-jet 
with a transverse momentum $p_T > 30$~GeV. The invariant mass of top 
candidates is shown in Fig.~\ref{topcandidates}. The estimate of the 
combinatorial background is shown as well. The top mass peak is clearly 
visible over the background. The top candidates with an invariant mass  
within $\pm 20$~GeV of the true top mass were selected.

The number of top candidates which passes these selections can be larger 
than two per event, 
resulting into a large number of possible top pairs combinations.
Because of the relativistic boost, the top pairs coming from the 
decay of one gluino have a smaller average angular separation than the 
pairs of top coming from the decay of different gluinos, or 
those coming from the Standard Model $t \bar t$ production. 
In order to reduce the Standard Model and the SUSY background, 
the angular separation $\Delta R$ (in the $(\eta,\phi)$ plane) 
between the two top candidates was required to be smaller 
than 2.5. If more than one top pair passes this selection, we select the
combination which minimizes

\begin{equation}
\sqrt{(m_{t1} - M_t)^2+(m_{t2} - M_t)^2}
\end{equation}

where $m_{t1}$ and $m_{t2}$ are the invariant masses of the two top candidates 
and $M_t$ is the true top mass.

\begin{table}[!tbp]
\begin{center}
\begin{tabular}{||c|c|c|c|c||} 
\hline 
Sample                    & Events      & inclusive cuts & two top \\
\hline 
SUSY signal               &        4708 &    597 &  51 \\   
SUSY back.                &       45292 &     15 &   1 \\
$t \bar t$                &  $7.6 \cdot 10^6$ &    397 &   3.3 \\
$W+jets$                  & $10.1 \cdot 10^6$ &     28 &   0.5 \\ 
$Z+jets$                  & $3.15 \cdot 10^6$ &     11 &   0.5 \\
$bb$+jets                 & $272 \cdot 10^6$  &    364 &   0   \\
\hline
\end{tabular}
\end{center}
\caption{\label{tab5} Efficiency of the cuts used for the reconstruction 
of the decay of the gluino into $t\bar t\tilde \chi^0$, 
evaluated with ATLFAST events for low luminosity operation. The number of 
events corresponds to an integrated luminosity of 10~$\mbox{fb}^{-1}$.
The third column contains the number of events which pass the 
inclusive cuts on jets, b-jets, missing energy and effective mass.  
The fourth column reports the number of events with two reconstructed  
top candidates which satisfy all cuts.  
SUSY events are divided in those with the 
presence of the $\tilde g \rightarrow \tilde \chi^0 t \bar t$  
decay (signal), and those without this decay (background).
}
\end{table}

The number of events which passes the various selections is shown in 
Table~\ref{tab5} for low-luminosity running conditions and 
an integrated luminosity of 10~$\mbox{fb}^{-1}$. The dominant
Standard Model backgrounds after the inclusive cuts on  
jets, b-jets, missing energy and effective mass (third column of 
Table~\ref{tab5}) are the $t\bar t$ and the $bb$+jets 
production. 
The latter is removed when the reconstruction 
of the hadronic decay of two top quarks with $\Delta R < 2.5$ is 
required (last column of the table), and the dominant background
remains the $t\bar t$ production, which is however more than one order of
magnitude smaller than the signal.

\begin{figure}[!htbp]
\begin{center}
\includegraphics[width=9cm]{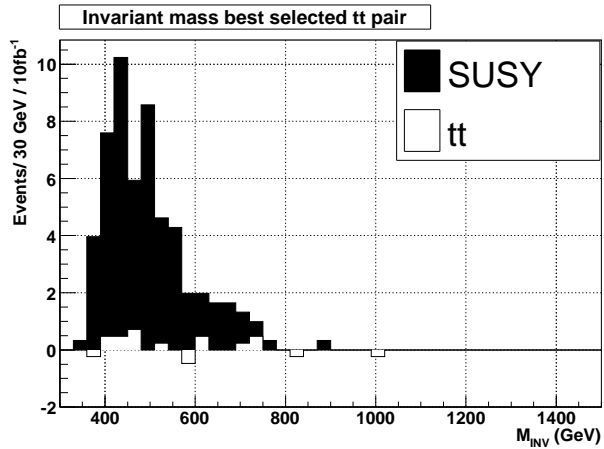}
\end{center} 
\caption{\label{tt1} Distribution of the invariant mass of the
selected pairs of reconstructed top quarks. The plot corresponds to  
10~$\mbox{fb}^{-1}$ of low-luminosity data.}
\end{figure}

The invariant mass of the top pair is shown in Fig.~\ref{tt1}
for low-luminosity running conditions and 
an integrated luminosity of 10~$\mbox{fb}^{-1}$. The 
statistical significance of the excess of events over the 
Standard Model Contribution is $\mbox{SUSY}/\sqrt{\mbox{SM}} = 7.1$
for an integrated luminosity of 1~$fb^{-1}$, 
which is comparable with the significance expected from the 
inclusive search and from the di-lepton analysis. 

\begin{figure}[!htbp]
\begin{center}
\includegraphics[width=9cm]{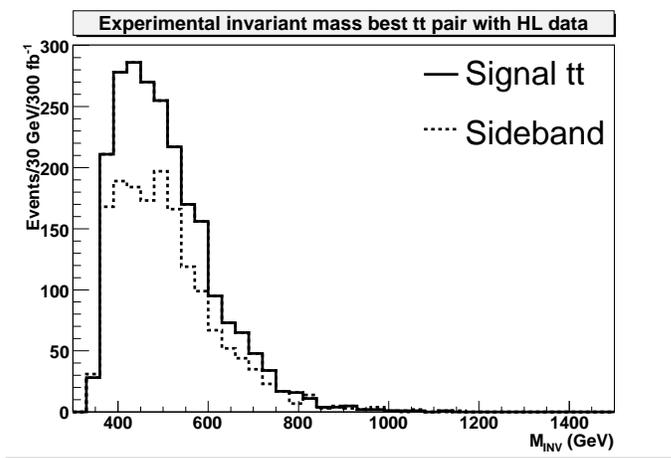}
\end{center} 
\caption{\label{tt2} Distribution of the invariant mass of the
selected pairs of reconstructed top quarks. The plot correspond to  
300~$\mbox{fb}^{-1}$ of high-luminosity data.}
\end{figure}

\begin{figure}[!htbp]
\begin{center}
\includegraphics[width=9cm]{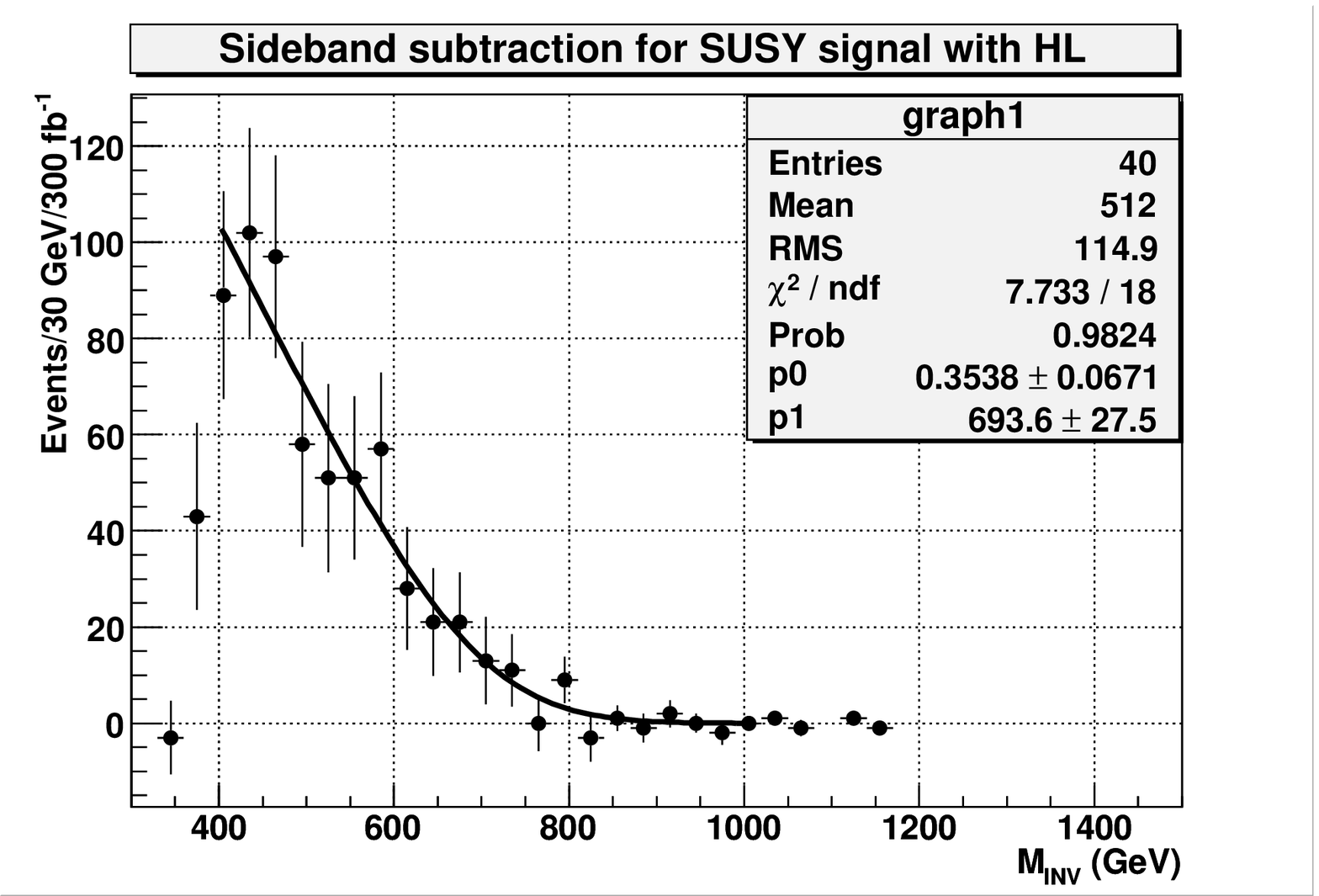}
\end{center} 
\caption{\label{tt3} Distribution of the invariant mass of the
selected pairs of reconstructed top quarks, after the subtraction of
the background estimated with the side-band technique.  
The plot correspond to 300~$\mbox{fb}^{-1}$ of high luminosity data.
}
\end{figure}

The distribution of $t\bar t$ 
pair invariant mass for high-luminosity conditions 
and an integrated statistic of $300~\mbox{fb}^{-1}$ is shown in 
Fig.~\ref{tt2}. The high luminosity implies a poorer jet resolution 
and a $b$-tagging efficiency of 0.5 for the same $u$-jet mistag probability 
of 0.01. Only the SUSY contribution is included in the analysis for 
high-luminosity. The 
contribution from the combinatorial background, estimated with 
top pairs built using the fake $W$ candidates, is also shown. 
The distribution obtained after the subtraction of this contribution 
is drawn in Fig.~\ref{tt3}. In order to estimate the position of 
the end point, a fit was performed with a polynomial of first order
convoluted with 
a Gaussian. The Gaussian represents the experimental resolution on 
the top pair invariant mass, and has been set to 15\% of the endpoint 
value. The agreement of this function with data is excellent. The  
endpoint is found to be $(694 \pm 28)$~GeV, in pretty good agreement with the 
value of $m(\tilde g) - m(\tilde \chi^0_2) = 696$~GeV. The interpretation of 
the endpoint in terms of this specific mass difference relies on the 
information from the Monte Carlo truth. Still, the endpoint of the
distribution of Fig.~\ref{tt3} provides an estimate of the mass
difference between the gluino and the neutralino states. 

\subsection{Reconstruction of the 
$\tilde g \rightarrow \tilde \chi^- t \bar b$ decay}

The reconstruction of the $t\bar b$ invariant mass using the decay 
$t\bar b \rightarrow jjb \bar b$ 
requires the presence of four jets, two of which 
tagged as b-jets to reduce the combinatorial background. 
Other jets are expected for signal events, from the decay of the other gluino
and/or the chargino. Thus, events were selected according to the 
following cuts:

\begin{itemize}

\item $E^T_{MISS} > 120$~GeV

\item At least one jet with $p_T > 200$~GeV, at least 6 jets 
with $p_T > 60$~GeV, at least two $b$-jets with  $p_T > 30$~GeV.

\item $M_{EFF} > 1200$~GeV


\end{itemize}

The top candidates were reconstructed as explained in the previous section. 
They were also required to have a transverse momentum $p_T > 150$~GeV. 
If more than one candidate satisfy these selections, the one with the  
closest invariant mass to the top mass was selected. 

Each of the top candidates was combined with the b-jets present in the event. 
Only the $t \bar b$ pairs with an angular separation $\Delta R < 2$ 
between the top reconstructed direction and the b-jet were 
considered. In the events where 
more than one $tb$ pair satisfy these selections, 
the smallest invariant mass of these pairs should be less than the 
kinematical endpoint of the signal.
Since combinatorial background pairs with 
an invariant mass larger than the kinematical endpoint of the signal 
would make the identification of the endpoint difficult, the combination with
the smallest invariant mass was selected. 

\begin{table}[!tbp]
\begin{center}
\begin{tabular}{||c|c|c|c|c||} 
\hline 
Sample          & Events & inclusive cuts & $t\bar b$ pair \\
\hline 
SUSY signal     &             3453  &    901 & 241 \\ 
SUSY background &            46547  &    313 &  91 \\
$t \bar t$      & $7.6  \cdot 10^6$ &   1127 & 143 \\
$W$+jets        & $10.1 \cdot 10^6$ &     60 &   5 \\
$Z$+jets        & $3.15 \cdot 10^6$ &     24 &   1 \\ 
$bb$+jets       & $272 \cdot 10^6$  &    873 &  41 \\
\hline
\end{tabular}
\end{center}
\caption{\label{tab6} Efficiency of the cuts used for the reconstruction 
of the decay of the gluino into $\tilde \chi^- t \bar b$, 
evaluated with ATLFAST events for low luminosity operation. The number of 
events corresponds to an integrated luminosity of 10~$\mbox{fb}^{-1}$.
The third column contains the number of events which pass the 
inclusive cuts on jets, b-jets, missing energy and effective mass.  
The fourth column reports the number of events with an accepted $t \bar b$  
pair. SUSY events are divided in events with the 
presence of the $\tilde g \rightarrow \tilde \chi^- t \bar b$  
decay in the Monte Carlo truth (signal), and without it (background). 
}
\end{table}

\begin{figure}[!htbp]
\begin{center}
\includegraphics[width=10cm]{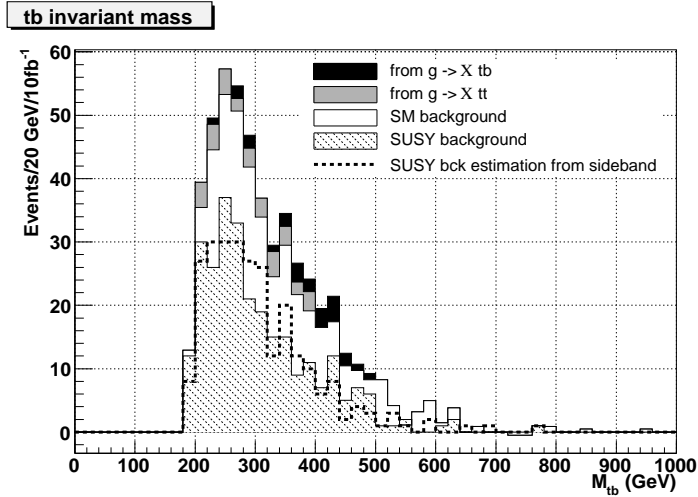}
\end{center} 
\caption{\label{tb} Distribution of the invariant mass of the
selected $t\bar b$ pairs. The plot correspond to  
10~$\mbox{fb}^{-1}$ of low-luminosity data.
}
\end{figure}

The efficiency of the various cuts is reported in Table~\ref{tab6}.
For an integrated luminosity of 10~$\mbox{fb}^{-1}$, 332 SUSY events 
and 190 Standard Model events pass all the selection cuts, including those
on the reconstructed $tb$ pair. In 241 
of the SUSY events the $\tilde g \rightarrow \tilde \chi^- t \bar b$  
decay is present in the Monte Carlo truth. They are classified as signal 
in the table, showing the efficiency of the selections for these events. 

However, the reconstructed $tb$ pair matches the one from the decay only in 
24 of those events. In the other SUSY events the reconstructed top 
does not have a correspondence in the MC truth (usually because of 
the incorrect pairing of jets to build the $W$ candidates), or the 
$tb$ pair matches a true top and bottom quarks originated from a 
$\tilde g \rightarrow \tilde \chi^0 t \bar t$ decay or 
from the decay of two different gluinos. The SUSY events thus 
pass the analysis selections with good efficiency, making this 
channel a promising one for SUSY discovery, but the probability to 
reconstruct the correct $tb$ pair is relatively small, making it 
difficult to reconstruct the kinematic endpoint in the invariant mass 
distribution.

The dominant Standard Model background is again $t\bar t$. The 
statistical significance $\mbox{SUSY}/\sqrt{\mbox{SM}}$ is 
7.6 for an integrated luminosity of 
1~$\mbox{fb}^{-1}$; this is comparable to the 
significance provided by the other searches presented here 
(inclusive, di-lepton 
and $t\bar t$ analysis). The  $\mbox{SUSY}/\mbox{SM}$ ratio is 
smaller, however, than in the leptonic and $t \bar t$ analysis, and the 
Standard Model contribution to the effective mass distribution 
cannot be neglected. The 
distribution of effective mass of the $t\bar b$ pairs is reported in 
Fig.~\ref{tb}, for an integrated luminosity 
of 10~$\mbox{fb}^{-1}$. The events are divided in the following classes:

\begin{itemize}

\item SUSY events in which the selected $t \bar b$ pair does 
indeed correspond to the pair coming from the decay 
$\tilde g \rightarrow \tilde \chi^{-} t \bar b$.

\item SUSY events, in which the reconstructed top and $b$-jet
does correspond to top and bottom quarks in the Monte Carlo truth,
but these quarks come from the decay 
$\tilde g \rightarrow \tilde \chi^0 t\bar t \rightarrow \tilde \chi^0 bW bW$. 
These 
events have not been classified under ``SUSY background'' because the 
invariant mass of the top with the bottom quark coming from the 
decay of the other top also has four  kinematic edges, corresponding to 
$m(\tilde g) - m(\tilde \chi^{0}) - m(W)$.

\item SUSY events in which the $b$-jet or the reconstructed top 
does not have a correspondence in the MC truth (usually because of 
the incorrect pairing of jets to build the $W$ candidates) 
or the corresponding quarks come from the decay chain of two different 
gluinos. These events are classified as ``SUSY background''.
In many of these events the  $\tilde g \rightarrow \tilde \chi^- t \bar b$ 
decay is present in the Monte Carlo truth, and are thus classified as signal 
in table~\ref{tab6}, but the reconstructed top and bottom do not 
match those from the decay. 
The contribution from the fake top candidates can be estimated from the 
sideband distribution reported as a dashed line.

\item Standard Model events

\end{itemize}

The contribution from the SUSY and $t \bar t$ backgrounds is large, and 
the statistics near the expected kinematic endpoint is scarce. It is 
thus not possible to measure the kinematic endpoint. 

Assuming that the contribution of the 
Standard Model background is reliably estimated, after subtraction of the 
Standard Model and combinatorial backgrounds it may be possible to obtain 
information on the $m(\tilde g) -m(\tilde \chi^{\pm})$ difference with 
larger statistics, but further study is required to evaluate this 
possibility. However, this analysis remains a promising strategy to find 
evidence of an excess of events over the Standard Model contribution.

\section{Extraction of the MSSM parameters}
\label{sec6}

In the MSSM, the neutralino and chargino sector of the theory 
depends on only four parameters: the gaugino masses $M_1$ and $M_2$, 
the Higgsino mass term $\mu$, and the ratio between the 
Higgs vacuum expectation values $\tan \beta$. 
In many models, including mSUGRA, the gaugino masses are unified at 
some high scale. This common value is the parameter $m_{1/2}$ in 
mSUGRA. At the TeV scale the relation 

\begin{equation}
\label{unif}
M_1 = \frac{5 g^{'2}}{3 g^2} M_2 \simeq 0.5 M_2
\end{equation}

holds, where $g$ and $g^{'}$ are the electroweak coupling constants.
This relation reduces the number of free parameters to three unknowns.

The constraints placed by the dilepton edge analysis 
on the neutralino mass spectrum can be used to
determine the values of $M_1$, $\mu$ and 
$\tan \beta$ which are compatible with the experimental measurements. 

The ISAJET~7.71 code~\cite{ISA} was used to compute the masses
and branching ratios of Supersymmetric particles as a function of 
the soft Supersymmetry breaking parameters. This provides, 
in particular, the values of neutralino and chargino masses 
as a function of  $M_1$, $M_2$, $\mu$ and $\tan \beta$.

For each point of parameter space the mass differences 
$\Delta M_2 = m(\tilde \chi^0_2)-m(\tilde \chi^0_1)$ and 
$\Delta M_3 = m(\tilde \chi^0_3)-m(\tilde \chi^0_1)$ were compared to the 
measured values $\Delta M_2^{\mbox{exp}} = 57.2$~GeV and 
$\Delta M_3^{\mbox{exp}} = 78.1$~GeV obtained using 
300~$\mbox{fb}^{-1}$ of simulated data, as discussed in Section~\ref{sec4}.
The following $\chi^2$ was evaluated: 

\begin{equation}
\chi^2 = (\Delta M_2 - \Delta M_2^{\mbox{exp}})^2/\sigma_2^2 
 + (\Delta M_3 - \Delta M_3^{\mbox{exp}})^2/\sigma_3^2 
 - 2r(\Delta M_2 - \Delta M_2^{\mbox{exp}})
       (\Delta M_3 - \Delta M_3^{\mbox{exp}})/\sigma_2 \sigma_3
\end{equation}
                                                                               
 where $\sigma_2 = 0.4$~GeV and $\sigma_3 = 1.4$~GeV are the errors on the 
measurements of $\Delta M_2$ and  $\Delta M_3$ respectively, and 
$r = 0.038$ is the correlation coefficient.

The points in the parameter space which have a $\chi^2$ probability 
larger than 0.05 were selected. In addition, the same sign was 
required for the $\tilde \chi^0_2$ and $\tilde \chi^0_1$ mass eigenstates 
and the opposite sign for the  $\tilde \chi^0_3$ and $\tilde \chi^0_1$
mass eigenstates, since a good fit of the di-lepton invariant mass 
distribution can be obtained only under these hypothesis, as 
discussed in Section~\ref{sec4}.

\begin{figure}[!htbp]
\begin{center}
\includegraphics[width=12cm]{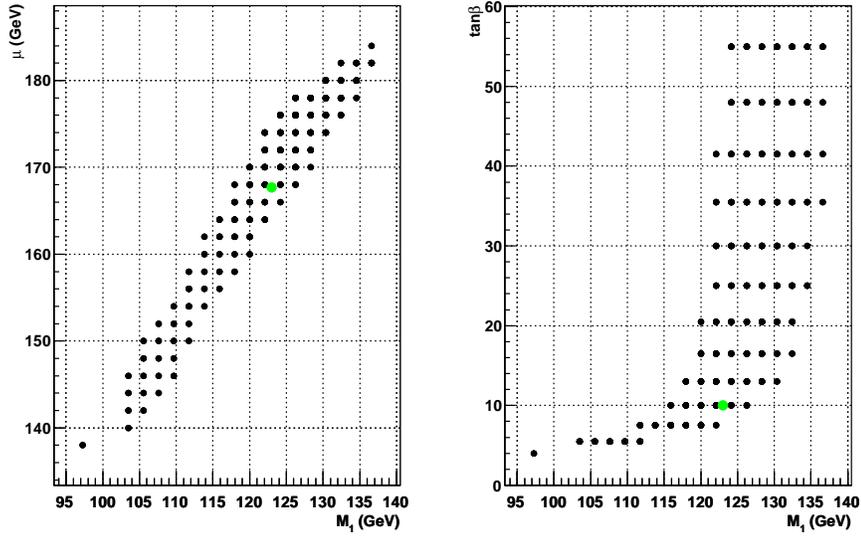}
\end{center} 
\caption{\label{parametri} Values of the MSSM parameters 
$\mu$, $M_1$ and $\tan \beta$ which are compatible with the 
experimental constraints on the neutralino mass spectrum 
obtained with 300~$\mbox{fb}^{-1}$ of data and at 95\% 
confidence level, for positive values of $\mu$.
}
\end{figure}

\begin{figure}[!htbp]
\begin{center}
\includegraphics[width=12cm]{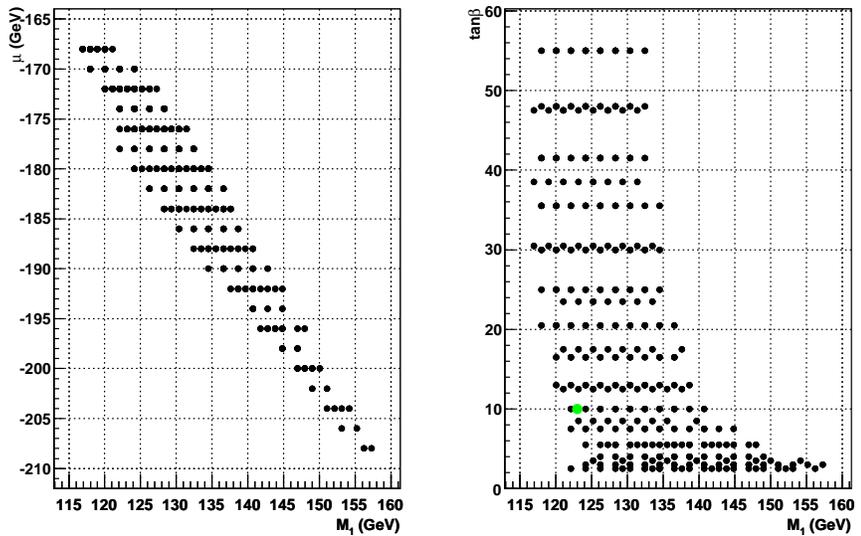}
\end{center} 
\caption{\label{parametri2} Values of the MSSM parameters 
$\mu$, $M_1$ and $\tan \beta$ which are compatible with the 
experimental constraints on the neutralino mass spectrum 
obtained with 300~$\mbox{fb}^{-1}$ of data and at 95\% 
confidence level, for negative values of $\mu$.
}
\end{figure}

A scan of the values of the parameters $M_1$, $\mu$ and 
$\tan \beta$ was then performed, using relation~\ref{unif} to fix the 
value of $M_2$, to find the parameter space which gives masses   
compatible with the simulated experimental constraints 
determined in section~\ref{sec4}. The scan was performed in the range 

\begin{equation}
45 \mbox{GeV} < M_1 <  500 \mbox{GeV}, \;\;\; 
45 \mbox{GeV} < |\mu| < 1000 \mbox{GeV}, \;\;\;
1 < \tan \beta < 65
\end{equation}

in steps of 2~GeV for $M_1$ and $\mu$, for 16 values of $tan \beta$,
and for both positive and negative values of $\mu$,  
so that a total of $3.4 \times 10^6$ values in the parameter space were 
considered. The region in this space which is compatible with the 
simulated ATLAS constraints is shown in Fig.~\ref{parametri} for 
positive values of $\mu$ and in Fig.~\ref{parametri2} for 
negative values of $\mu$.
 
The upper left plot shows the allowed values of $M_1$ and $\mu$. 
Each parameter can vary in a relatively large interval: 
$97 \; \mbox{GeV} < M_1 < 136$~GeV and $138 \; \mbox{GeV} < \mu < 182$~GeV
for positive values of $\mu$, $116 \; \mbox{GeV} < M_1 < 157$~GeV 
and $-208 \; \mbox{GeV} < \mu < -168$~GeV for negative values of 
$\mu$. However, the ratio is more constrained 
to the range $1.32 < \mu/M_1 < 1.46$. 
The value of $\tan \beta$ is not constrained by the experimental data.

The value of the ratio between $\mu$ and $M_1$ is interesting since in 
most of mSUGRA space $\mu >> M_2 \simeq 2 M_1$. With this hierarchy between 
the parameters, the lightest neutralino is almost a pure Bino. 
The focus point region of mSUGRA, instead, is characterized by 
$\mu \simeq M_1$~\cite{Focuspoint}. In this case the mixing between gauge 
eigenstates is nearly maximal and each mass eigenstate receives a significant 
contribution from all the gauge eigenstates. The Higgsino component 
of the $\tilde \chi^0_1$ allows rapid s-channel neutralino annihilation, 
which is the mechanism which reduces the relic density in the early 
universe. The measurement of the two leptonic edges allows to establish this 
scenario.

For any given set of values of the  parameters $M_1, M_2, \mu,$ 
and $\tan \beta$ the masses of the neutralinos and their gaugino 
mixing angles can be computed, so these quantities are also 
constrained by the data. For the values of the 
parameters which are compatible with the dilepton invariant mass data 
at 95\% C.L. the mass of the lightest neutralino 
lies in the range $67 \; \mbox{GeV} < m(\tilde \chi^0_1) < 156$~GeV.
For any given value of the mass of $\tilde \chi^0_1$, the mass of the next two 
lightest neutralinos is precisely constrained from the measurements of the 
two edges of the dilepton distribution.

\section{Conclusions}

A study of the ATLAS potential of detecting and measuring 
Supersymmetry in the 
focus point scenario has been presented. 
For the selected point in the parameter space the 
observation of an excess of events over the Standard Model expectations 
should be observed rather easily; the time needed for discovery would 
probably be determined by the understanding of the systematics related to 
the detector response and the knowledge of the Standard Model backgrounds 
rather than the statistical significance. Several channels can contribute to
the discovery of an excess of SUSY events with a comparable statistical 
significance: with appropriate kinematic cuts, the presence of 
Supersymmetry physics may become manifest through an 
excess of events with hard jets, large missing energy 
and $b$-jets (section~\ref{sec3}), events with hard jets, large missing 
energy and opposite-sign 
electron or muon pairs (section~\ref{sec4}), and 
events with hard jets, large missing 
energy, a top and a bottom quark or two top quarks 
(section~\ref{sec5}). In each of these channels, the
contribution from SUSY events has a statistical significance between 
6.1 and 8.2 standard deviations with 1~$\mbox{fb}^{-1}$ of data.

In the focus point region, the neutralino leptonic decays proceed 
through a direct three-body decay in which the virtual slepton exchange
is negligible, because of the large scalar mass. 
An analytical expression was derived in section~\ref{sec4} for the 
distribution of the invariant mass of the resulting lepton pairs.
This formula was used to fit the distribution of opposite-sign 
lepton pairs obtained with 300~$\mbox{fb}^{-1}$ of simulated LHC data. 
Two kinematic edges, measuring the  
$m(\tilde \chi^0_3) - m(\tilde \chi^0_1)$ and
$m(\tilde \chi^0_2) - m(\tilde \chi^0_1)$ mass differences, are 
measured with a precision of the order of 1~GeV. 

The constraints that this measurement would place on the MSSM gaugino 
sector parameters have been discussed. These constraints are such that    
a scenario with a large mixing in the neutralino sector and a relic 
neutralino density of the same order of magnitude as the Dark Matter 
abundance would emerge from the LHC measurements.

The gluino decay into $t \bar t\tilde \chi^0$ and $t \bar b\tilde \chi^+$ 
can be studied 
through the reconstruction of the $tt$ and $t \bar b$ invariant mass. 
While a precise reconstruction of the corresponding kinematic edges 
does not seem to be possible, evidence for these decays and an 
estimate of the mass difference between the gluino and gaugino states 
may be extracted from the experimental data.

\newpage

\begin{center}
{\bf Acknowledgments}
\end{center}

The authors wish to thank G. Polesello (INFN Pavia) for his helpful 
suggestions on the analysis and on the preparation of this document. 
They also thank D. Tovey (Sheffield), F. Paige (BNL) and 
I. Hinchliffe (LBNL), as well as M.M. Nojiri (KEK, Tsukuba),
for useful discussions and advices.

The authors are also grateful to S. Asai and the colleagues of 
the University of Tokyo for providing 
the ALPGEN 4-vector files used in the production of the   
W+jets and Z+jets backgrounds, 
and to B. Gjelsten (Bern) for the production of the Z+jets fast simulation 
ntuples.

This work has been performed within the ATLAS collaboration 
making use of 
the detector fast simulation and tools which are the result of 
collaboration-wide efforts.



\begin{thebibliography}{99}

\bibitem{SUSY} See H.P. Nilles, Phys. Rev. 110 (1984) 1 
and references therein. \\

\bibitem{LEP} See the LEP (ALEPH+DELPHI+L3+OPAL experiments) SUSY 
working group page, http://lepsusy.web.cern.ch/lepsusy, and references 
therein. \\

\bibitem{WMAP} D.N. Spergel et al., astro-ph/0603449. \\ 
C.L. Bennett et al., Astrophys. J. Suppl. 148 (2003) 1. \\
D.N.Spergel et al., Astrophys. J. Suppl. 148 (2003) 175. \\

\bibitem{Ell03} J.R. Ellis et al., Phys. Lett. B565 (2003) 176. \\
H.W. Baer and C. Balazs, J. Cosmol. Astr, Phys. 0305 (2003) 006. \\

\bibitem{Focuspoint}
J.L. Feng, K.T. Matchev, and T. Moroi, Phys. Rev. Lett. 84 (2000) 2322 
[hep-ph/9908309]. \\
J.L. Feng, K.T. Matchev, and T. Moroi, Phys. Rev. D61 (2000) 075005
[hep-ph/9909334]. \\
J.L. Feng, K.T. Matchev, and F. Wilczek, Phys.Lett.B482 (2000) 388 
[hep-ph/0004043]. \\ 
J.L. Feng, K.T. Matchev, and F. Wilczek, Phys. Rev. D63 (2001) 045024 
[hep-ph/0008115]. \\

\bibitem{mOMEGAs}  G. B. Belanger, F. Boudjema, A. Pukhov and A. Semenov,  
Comput. Phys. Commun. 174 (2006) 577  [hep-ph/0405253]. \\

\bibitem{ISA}  H. Baer, F.E. Paige, S.D. Prototopescu and X. Tata, 
hep-ph/0312045. \\

\bibitem{SOFT}  B.C. Allanach, Comput. Phys. Commun. 143 (2002) 305-331, 
hep-ph/0104145. \\

\bibitem{All03} B.C. Allanach et al., JHEP 0303 (2003) 016
[hep-ph/0302102] and references therein. \\

\bibitem{All04} B.C. Allanach et al., hep-ph/0402161. \\
G. Belanger, S. Kraml, and A. Pukhov, 
Phys. Rev. D72 (2005) 015003 [hep-ph/0502079]. \\

\bibitem{ATDR} ATLAS Physics Performance and Detector Technical Design Report,
CERN/LHCC 99-14 (1999). \\

\bibitem{TevTop} Tevatron electroweak working group, hep-ex/0608032.

\bibitem{HERWIG} G. Corcella et al., JHEP 0101 (2001) 010. \\
S. Moretti et al., JHEP 0204 (2002) 028. \\
G. Corcella et al., hep-ph/02010213. \\

\bibitem{MCNLO} S. Frixione and B.R. Webber, 
JHEP 0206 (2002) 029 [hep-ph/0204244]. \\
S. Frixione, P. Nason and B.R. Webber, 
JHEP 0308 (2003) 007 [hep-ph/0305252]. \\


\bibitem{ALPGEN} M.L. Mangano, M. Moretti, F. Piccinini, R. Pittau, A. Polosa, 
JHEP 0307 (2003) 001 [hep-ph/0206293]. \\
M.L. Mangano, M. Moretti, R. Pittau, 
Nucl.Phys. B632 (2002) 343 [hep-ph/0108069]. \\
F. Caravaglios, M. L. Mangano, M. Moretti, R. Pittau, 
Nucl.Phys.B539 (1999) 215 [hep-ph/9807570]. \\

\bibitem{ATLFAST} E. Richter-Was, D. Froidevaux and L. Poggioli, 
ATL-PHYS-96-079 and ATL-PHYS-98-131. \\

\bibitem{trigTDR} ATLAS High-Level Trigger, Data Acquisition and Controls 
Technical Design Report, CERN/LHCC 2003-022 (2003).

\bibitem{Mih99} M.M. Nojiri and Y. Yamada, Phys. Rev. D60 (1999) 015006. \\

\bibitem{Mon05} S. Montesano, ``Ricerca di particelle supersimmetriche 
nell'ambito dell'esperimento ATLAS'', diploma thesis, University of Milano
(2006).

\bibitem{His03} J. Hisano, K. Kawagoe and M. Nojiri, Phys. Rev. D68 (2003) 
035007. \\

\bibitem{Gje02} C.K. Gjelsten et al., ATL-PHYS-2004-007. \\
C.K. Gjelsten et al., JHEP 12 (2004) 003. \\

\end{thebibliography}
\end{document}